\documentclass{article}
\usepackage{eurosym}
\usepackage{amssymb}
\usepackage{amsmath}
\usepackage{cite}
\usepackage{graphicx}
\usepackage{epsf}
\usepackage{lscape}

\begin{document}

\title{Exact Anti-Self-Dual four-manifolds with a Killing symmetry by
similarity transformations}
\author{Andronikos Paliathanasis\thanks{%
Email: anpaliat@phys.uoa.gr} \\
{\ \textit{Institute of Systems Science, Durban University of Technology }}\\
{\ \textit{PO Box 1334, Durban 4000, Republic of South Africa}\ } \\
{\textit{Instituto de Ciencias F\'{\i}sicas y Matem\'{a}ticas,}}\\
{\ \textit{Universidad Austral de Chile, Valdivia, Chile}}}
\maketitle

\begin{abstract}
We study the group properties and the similarity solutions for the
constraint conditions of anti-self-dual null K\"{a}hler four-dimensional
manifolds with at least a Killing symmetry vector. Specifically we apply the
theory of Lie symmetries to determine all the infinitesimal generators of
the one-parameter point transformations which leave the system invariant. We
use these transformations to define invariant similarity transformations
which are used to simplify the differential equations and find the exact
form of the spacetime. We show that the constraint equations admit an
infinite number of symmetries which can be used to construct an infinite
number of similarity transformations.

\bigskip

\end{abstract}

\section{Introduction}

\label{sec1}

By definition, an Anti-Self-Dual Riemannian space has the property its self
dual Weyl curvature to be zero, Locally conformally flat spacetimes are
known examples of ASD manifolds. The property a manifold is ASD is usually
related with the context of integrability for specific field equations of
Riemannian manifolds. The pioneer work of Penrose relates four-dimensional
ASD manifolds with three-dimensional complex twistor spaces with certain
algebraic properties \cite{pen1}. Furthermore, Ward \cite{ward} generalized
the result of Penrose by introducing the cosmological constant in the field
equations. There are various studies in the literature on the investigation
of ASD four-manifolds. For instance, in \cite{as1}, Gibbons and Hawking
found a new family of class of hyper--K\"{a}hler metrics characterized by
the existence of an isometry. LeBrun \cite{as2} generalized that result; he
proved that self-dual four-manifolds with $G_{1}$ symmetry can be
constructed from the solutions of a linear differential equation on a
three-dimensional hyperbolic space. Such approach was applied in \cite{as3}
to determine self-dual manifolds of higher dimension with a $G_{n}$ symmetry
group.

In this study we focus on ASD null K\"{a}hler metrics of signature $\left(
++--\right) $ which admit a covariantly constant real spinor and an isometry
vector field. Hence, we assume the line element
\begin{equation}
ds^{2}=W_{x}\left( dy^{2}-4dxdt+4H_{x}dt^{2}\right) -W_{x}^{-1}\left(
dz-W_{x}dy-2W_{y}dt\right) ^{2}  \label{as.03}
\end{equation}%
which we shall say that it is ASD null K\"{a}hler geometry if and only if
functions $H\left( t,x,y\right) $ and $W\left( t,x,y\right) $ satisfy the
following system \cite{asd1}
\begin{eqnarray}
H_{yy}-H_{tx}+H_{x}H_{xx} &=&0,  \label{as.01} \\
W_{yy}-W_{tx}+\left( H_{x}W_{x}\right) _{,x} &=&0,  \label{as.02}
\end{eqnarray}%
where the admitted Killing symmetry is the vector field $K=\partial _{z}$,
that is, $\mathcal{L}_{K}g_{\mu \nu }=0$, where $\mathcal{L}_{K}$ is the Lie
derivative with respect to the vector field $K$ on the metric tensor $g_{\mu
\nu }$ with line element $ds^{2}=g_{\mu \nu }dx^{\mu }dx^{\nu }$ as defined
by expression (\ref{as.03}) and $x^{\mu }=\left( t,x,y,z\right) $. We
observe that for $W=\frac{H_{,x}}{2}+f\left( t\right) $ the line element
describes a pseudo hyper-K\"{a}hler metric, while when $W_{x}\neq \frac{%
H_{xx}}{2}$ the line element describes a non Ricci-flat space. For $H\left(
t,x,y\right) =H_{0}$, the space reduces to the Gibbons-Hawking solution \cite%
{asd1}.

We are interested on the algebraic properties of system (\ref{as.01}), (\ref%
{as.02}) and in the derivation of exact solutions. Specifically, we consider
the theory of the symmetries of differential equations and we determine all
the Lie point symmetries for the gravitational system (\ref{as.01}), (\ref%
{as.02}). Lie symmetry analysis is a robust method for the study of
nonlinear differential equations. Nowadays it is the standard approach for
the computation of solutions and the description of the algebra for
nonlinear differential equations \cite{ovsi,ibra,Bluman,olver}, for
applications see \cite{l1,l2,l3,l4,l5,l6,l7,l8,l9,l10,l11,l12,l13,l14,l15}
and references therein. In this work we take some interest in the
application of Lie symmetries in the classification of exact solutions. This
specific approach has been widely studied in various gravitational systems
from General Relativity to alternative theories of gravity with very
interesting results \cite{gr1,gr2,gr3,gr4,gr5}. For a recent review on the
application of Lie symmetries in cosmological studies we refer the reader to
\cite{gr6}. Indeed, as we shall see in the following, with the application
of the Lie point symmetries we find new families of solutions for the system
(\ref{as.01}), (\ref{as.02}). We recover previous well-known solutions, thus
we find new solutions which have not been presented before in the
literature. Indeed for the plethora of these solutions the resulting
spacetime admits additional Killing vectors. However, there are exact
solutions where no additional Killing vectors exists for the line element (%
\ref{as.03}).

As far as the Kadomtsev--Petviashvili equation is concerned, the group
properties have been investigated in various studies in the literature \cite%
{kp1,kp2,kp3}. Kadomtsev--Petviashvili equation is related with the
Einstein-Weyl equations in three-dimensional manifolds. Indeed, if a
three-dimensional\ Einstein-Weyl structure admits an isometry then its
structure is defined by the Kadomtsev--Petviashvili equation \cite{a12}. The
plan of the paper is as follows.

In Section \ref{sec2} we present the basic properties and definitions for
Lie symmetries and the main definition of the one-dimensional optimal
system. We continue our analysis by investigating two separate cases; the
spacetime (\ref{as.03}) being static or non static. The Lie symmetries of
the field equations (\ref{as.01}), (\ref{as.02}) \ for the static spacetime
are investigated in Section \ref{sec3}. We find that the system admits a
nine-dimensional Lie algebra. We calculate the commutators and the adjoint
representation, the latter results are used to define the one-dimensional
optimal system which for simplicity of the presentation we give it terms of
various families. The Lie point symmetries are applied to define similarity
transformations which are used to reduce the gravitational field equations
into a system of ordinary differential equations. For all the reductions of
our analysis the reduced system can be solved in terms of quadratures.
However, in this work we investigate the possibility of the exact
closed-form solutions. For the exact spacetimes we study the admitted
isometries which can be used for the physical description of the solutions.
The nonstatic case is investigated in Section \ref{sec4}. In contrast with
the static case, we find that the gravitational field equations admit
infinite Lie point symmetries, which are summarized in nine families of Lie
vector fields. These families provide eight coefficient functions which
depend on the time parameter. We demonstrate the application of these
infinite Lie symmetries in the derivation of nonstatic exact closed-form
solutions. Finally in Section \ref{sec5}, we summarize our results and we
draw our conclusions.

\section{Lie symmetries and one-dimensional optimal system}

\label{sec2}

For the convenience of the reader in the following lines we present the
basic properties and definitions for the theory of Lie symmetries.

Let us consider the infinitesimal one-parameter point transformation%
\begin{eqnarray}
t^{\prime } &=&t+\varepsilon \xi ^{t}\left( t,x,y,H,W\right) ,  \label{as.04}
\\
x^{\prime } &=&x+\varepsilon \xi ^{x}\left( t,x,y,H,W\right) ,  \label{as.05}
\\
y^{\prime } &=&y+\varepsilon \xi ^{y}\left( t,x,y,H,W\right)  \label{as.06}
\\
H^{\prime } &=&H+\varepsilon \eta ^{H}\left( t,x,y,H,W\right)  \label{as.07}
\\
W^{\prime } &=&W+\varepsilon \eta ^{W}\left( t,x,y,H,W\right)  \label{as.08}
\end{eqnarray}%
with infinitesimal generator $X=\xi ^{\mu }\partial _{\mu }+\eta
^{A}\partial _{A}$,~$\xi ^{\mu }=\left( \xi ^{t},\xi ^{x},\xi ^{y}\right) $,
$\eta ^{A}=\left( \eta ^{H},\eta ^{W}\right) $ \ and second
extension/prolongation
\begin{equation}
X^{\left[ 2\right] }=\xi ^{\mu }\partial _{\mu }+\eta ^{A}\partial _{A}+\eta
^{A\left[ 1\right] }\partial _{A_{,}\mu }+\xi ^{\mu }\partial _{\mu }+\eta
^{A\left[ 2\right] }\partial _{A_{,\mu \nu }}.  \label{as.09}
\end{equation}%
in which the coefficients $\eta ^{A\left[ 1\right] },~\eta ^{A\left[ 2\right]
}$ are defined by the following formula
\begin{equation}
\eta ^{\left[ n\right] }=D_{\mu }\eta ^{\left[ n-1\right] }-u_{\mu _{1}\mu
_{2},...,\mu _{n-1}}D_{\mu }\left( \xi ^{\mu }\right) .  \label{as.10}
\end{equation}

Therefore, under the action of the latter point transformation the
gravitational system (\ref{as.01}), (\ref{as.02}) remain invariant if and
only if \cite{ibra,Bluman,olver}%
\begin{equation}
\lim_{\varepsilon \rightarrow 0}\frac{\mathbf{F}\left( t,x,y,H,W;\varepsilon
\right) -\mathbf{F}\left( t,x,y,H,W;\varepsilon \right) }{\varepsilon }=0,
\label{as.11}
\end{equation}%
where $\mathbf{F}=\left( H_{yy}-H_{tx}+H_{x}H_{xx},W_{yy}-W_{tx}+\left(
H_{x}W_{x}\right) _{,x}\right) $. The symmetry condition (\ref{as.11}) can
be written in the equivalent form%
\begin{equation}
\mathcal{L}_{X^{\left[ 2\right] }}\left( \mathbf{F}\right) =\lambda \mathbf{%
F~,~}{mod~}\mathbf{F}=0  \label{as.12}
\end{equation}%
where $\lambda $ is a function that should be determined and $L_{X^{\left[ 2%
\right] }}$ denotes the Lie derivative with respect to the vector field $X^{%
\left[ 2\right] }$.

The main application of the Lie symmetries is summarized in the derivation
of similarity transformations which are used to simplify the differential
equation by reduce the number of the independent variables or by reducing
the order of the differential equation. The final purpose for the
application of the Lie symmetries is to reduce the given differential
equation into the form of another well-known equation or write the
differential equation into a simple form where an exact solution can be
found.

The solutions which follow from the application of the Lie invariants are
mainly known as similarity solutions. A basic property for the admitted Lie
symmetries of a differential equation is that they form a Lie group.
Therefore, in order to perform a complete derivation of all the possible
similarity solutions we should find the admitted one-dimensional optimal
system.

Consider the $n$-dimensional Lie algebra $G_{n}$\ with elements $%
X_{1},~X_{2},~...~X_{n}$\ admitted by the system $H^{A}$. Then the vector
fields \cite{ovsi,olver}
\begin{equation}
Z=\sum\limits_{i=1}^{n}a_{i}X_{i}~,~W=\sum\limits_{i=1}^{n}b_{i}X_{i}~,~%
\text{\ }a_{i},~b_{i}\text{ are constants.}  \label{as.13}
\end{equation}%
are equivalent if and only if%
\begin{equation}
\mathbf{W}=Ad\left( \exp \left( \varepsilon _{i}X_{i}\right) \right) \mathbf{%
Z}  \label{as.14}
\end{equation}%
or%
\begin{equation}
W=cZ~,~c=const.  \label{as.15}
\end{equation}%
where the operator $Ad\left( \exp \left( \varepsilon X_{i}\right) \right)
X_{j}=X_{j}-\varepsilon \left[ X_{i},X_{j}\right] +\frac{1}{2}\varepsilon
^{2}\left[ X_{i},\left[ X_{i},X_{j}\right] \right] +...~$is called the
adjoint representation. It is clear that the adjoint representation of the
admitted Lie symmetries should be calculated in order to find all the
possible independent similarity transformations. The latter set of
one-dimensional Lie algebras which do not connect thought the adjoint
representation form the so-called one-dimensional optimal system for the
given differential equation.

We proceed our analysis by considering first static spacetime with $%
H=H\left( x,y\right) ,~W=W\left( x,y\right) $\ and then the general case. We
follow that analysis for the convenience of the reader and the presentation
of the results.

\section{Static spacetime}

\label{sec3}

In the case where the spacetime is static (\ref{as.03}), that is, it admits
the second isometry vector $\partial _{t}$, the gravitational system (\ref%
{as.01}), (\ref{as.02}) is simplified as%
\begin{eqnarray}
H_{yy}+H_{x}H_{xx} &=&0,  \label{as.16} \\
W_{yy}+\left( H_{x}W_{x}\right) _{,x} &=&0.  \label{as.17}
\end{eqnarray}

From the symmetry condition (\ref{as.12}) for the latter system we find the
following symmetry vectors%
\begin{eqnarray*}
X^{1} &=&\partial _{H}~,~X^{2}=\partial _{W}~,~X^{3}=\partial
_{x}~,~X^{4}=\partial _{y}~,~X^{5}=W\partial _{W}~, \\
X^{6} &=&y\partial _{H}~,~X^{7}=y\partial _{W}~,~X^{8}=H\partial _{H}-\frac{y%
}{2}\partial _{y}~,~X^{9}=x\partial _{x}+\frac{3}{2}y\partial _{y}.
\end{eqnarray*}%
The corresponding commutators and the adjoint representation for the
admitted Lie symmetries are presented in Tables \ref{tab1} and \ref{tab2}
respectively. With the results of the tables we can determine the
one-dimensional optimal system, consisting of the following one-dimensional
Lie algebras%
\begin{eqnarray*}
&&\left\{ X^{1}\right\} ~,~\left\{ X^{2}\right\} ~,~\left\{ X^{3}\right\}
~,~\left\{ X^{4}\right\} ~,~\left\{ X^{5}\right\} ~,~\left\{ X^{6}\right\}
~,~\left\{ X^{7}\right\} ~,~\left\{ X^{8}\right\} ~,~\left\{ X^{9}\right\} ~,
\\
&&\left\{ a_{2}X_{2}+a_{8}X_{8}+a_{9}X_{9}\right\} ~,~\left\{
a_{3}X_{3}+a_{5}X_{5}+a_{8}X_{8}\right\} ~,~\left\{
a_{1}X_{1}+a_{5}X_{5}+a_{9}X_{9}\right\} ~ \\
&&\left\{ a_{1}X_{1}+a_{2}X_{2}+a_{9}X_{9}\right\} ~,~\left\{
a_{2}X_{2}+a_{3}X_{3}+a_{8}X_{8}\right\} ~,~\left\{
a_{3}X_{3}+a_{4}X_{4}+a_{5}X_{5}\right\} ~,~ \\
&&\left\{ a_{3}X_{3}+a_{4}X_{4}+a_{6}X_{6}+a_{7}X_{7}\right\} ~,~\left\{
a_{3}X_{3}+a_{4}X_{4}+a_{5}X_{5}+a_{6}X_{6}\right\} ~,~ \\
&&\left\{ a_{2}X_{2}+a_{3}X_{3}+a_{4}X_{4}+a_{6}X_{6}\right\} ~,~\left\{
a_{1}X_{1}+a_{2}X_{2}+a_{3}X_{3}+a_{6}X_{6}\right\} ~, \\
&&\left\{ a_{1}X_{1}+a_{3}X_{3}+a_{4}X_{4}+a_{5}X_{5}\right\} ~,~\left\{
a_{1}X+a_{2}X_{2}+a_{3}X_{3}+a_{4}X_{4}\right\} ~
\end{eqnarray*}%
in which $a_{I}$ are real numbers. \ We have used the coefficients $a_{I}$
such as to simplify the presentation of the one-dimensional optimal system.

We proceed with the application of the Lie point symmetries for the
derivation of the similarity transformations which will be used to reduce
the number of the independent variables of the gravitational system (\ref%
{as.15}), (\ref{as.17}). The system (\ref{as.15}), (\ref{as.17}) is reduced
into a system of ordinary differential equation which as we shall see in all
cases can be solved by quadratures.

\begin{table}[tbp] \centering%
\caption{Commutator table for the Lie point symmetries of the statis
gravitational system (\ref{as.16})-(\ref{as.17}).}%
\begin{tabular}{cccccccccc}
\hline\hline
$\left[ X_{I},X_{J}\right] $ & $\mathbf{X}_{1}$ & $\mathbf{X}_{2}$ & $%
\mathbf{X}_{3}$ & $\mathbf{X}_{4}$ & $\mathbf{X}_{5}$ & $\mathbf{X}_{6}$ & $%
\mathbf{X}_{7}$ & $\mathbf{X}_{8}$ & $\mathbf{X}_{8}$ \\ \hline
$\mathbf{X}_{1}$ & $0$ & $0$ & $0$ & $0$ & $0$ & $0$ & $0$ & $X_{1}$ & $0$
\\
$\mathbf{X}_{2}$ & $0$ & $0$ & $0$ & $0$ & $X_{2}$ & $0$ & $0$ & $0$ & $0$
\\
$\mathbf{X}_{3}$ & $0$ & $0$ & $0$ & $0$ & $0$ & $0$ & $0$ & $0$ & $X_{3}$
\\
$\mathbf{X}_{4}$ & $0$ & $0$ & $0$ & $0$ & $0$ & $X_{1}$ & $X_{2}$ & $-\frac{%
1}{2}X_{4}$ & $\frac{3}{2}X_{4}$ \\
$\mathbf{X}_{5}$ & $0$ & $-X_{2}$ & $0$ & $0$ & $0$ & $0$ & $-X_{7}$ & $0$ &
$0$ \\
$\mathbf{X}_{6}$ & $0$ & $0$ & $0$ & $-X_{1}$ & $0$ & $0$ & $0$ & $\frac{3}{2%
}X_{6}$ & $-\frac{3}{2}X_{6}$ \\
$\mathbf{X}_{7}$ & $0$ & $0$ & $0$ & $-X_{2}$ & $X_{7}$ & $0$ & $0$ & $\frac{%
1}{2}X_{7}$ & $-\frac{3}{2}X_{7}$ \\
$\mathbf{X}_{8}$ & $-X_{1}$ & $0$ & $0$ & $\frac{1}{2}X_{4}$ & $0$ & $-\frac{%
3}{2}X_{6}$ & $-\frac{1}{2}X_{7}$ & $0$ & $0$ \\
$\mathbf{X}_{9}$ & $0$ & $0$ & $-X_{3}$ & $\frac{3}{2}X_{4}$ & $0$ & $\frac{3%
}{2}X_{6}$ & $\frac{3}{2}X_{7}$ & $0$ & $0$ \\ \hline\hline
\end{tabular}%
\label{tab1}%
\end{table}%

\begin{table}[tbp] \centering%
\caption{Adjoint representation for the Lie point symmetries of the statis
gravitational system (\ref{as.16})-(\ref{as.17}).}%
\begin{tabular}{cccccccccc}
\hline\hline
$Ad\left( e^{\left( \varepsilon \mathbf{X}_{i}\right) }\right) \mathbf{X}%
_{j} $ & $\mathbf{X}_{1}$ & $\mathbf{X}_{2}$ & $\mathbf{X}_{3}$ & $\mathbf{X}%
_{4}$ & $\mathbf{X}_{5}$ & $\mathbf{X}_{6}$ & $\mathbf{X}_{7}$ & $\mathbf{X}%
_{8}$ & $\mathbf{X}_{8}$ \\ \hline
$\mathbf{X}_{1}$ & $X_{1}$ & $X_{2}$ & $X_{3}$ & $X_{4}$ & $X_{5}$ & $X_{6}$
& $X_{7}$ & $X_{8}-\varepsilon X_{1}$ & $X_{9}$ \\
$\mathbf{X}_{2}$ & $X_{1}$ & $X_{2}$ & $X_{3}$ & $X_{4}$ & $%
X_{5}-\varepsilon X_{2}$ & $X_{6}$ & $X_{7}$ & $X_{8}$ & $X_{9}$ \\
$\mathbf{X}_{3}$ & $X_{1}$ & $X_{2}$ & $X_{3}$ & $X_{4}$ & $X_{5}$ & $X_{6}$
& $X_{7}$ & $X_{8}$ & $X_{9}-\varepsilon X_{3}$ \\
$\mathbf{X}_{4}$ & $X_{1}$ & $X_{2}$ & $X_{3}$ & $X_{4}$ & $X_{5}$ & $%
X_{6}-\varepsilon X_{1}$ & $X_{7}-\varepsilon X_{1}$ & $X_{8}+\frac{%
\varepsilon }{2}X_{4}$ & $X_{9}-\frac{3}{2}\varepsilon X_{4}$ \\
$\mathbf{X}_{5}$ & $X_{1}$ & $e^{\varepsilon }X_{2}$ & $X_{3}$ & $X_{4}$ & $%
X_{5}$ & $X_{6}$ & $e^{\varepsilon }X_{7}$ & $X_{8}$ & $X_{9}$ \\
$\mathbf{X}_{6}$ & $X_{1}$ & $X_{2}$ & $X_{3}$ & $X_{4}+\varepsilon X_{1}$ &
$X_{5}$ & $X_{6}$ & $X_{7}$ & $X_{8}-\frac{3}{2}\varepsilon X_{6}$ & $X_{9}+%
\frac{3}{2}\varepsilon X_{6}$ \\
$\mathbf{X}_{7}$ & $X_{1}$ & $X_{2}$ & $X_{3}$ & $X_{4}+\varepsilon X_{2}$ &
$X_{5}-\varepsilon X_{7}$ & $X_{6}$ & $X_{7}$ & $X_{8}-\frac{1}{2}%
\varepsilon X_{7}$ & $X_{9}+\frac{3}{2}\varepsilon X_{7}$ \\
$\mathbf{X}_{8}$ & $e^{\varepsilon }X_{1}$ & $X_{2}$ & $X_{3}$ & $e^{-\frac{%
\varepsilon }{2}}X_{4}$ & $X_{5}$ & $e^{\frac{3}{2}\varepsilon }X_{6}$ & $e^{%
\frac{\varepsilon }{2}}X_{7}$ & $X_{8}$ & $X_{9}$ \\
$\mathbf{X}_{9}$ & $X_{1}$ & $X_{2}$ & $e^{\varepsilon }X_{3}$ & $e^{\frac{3%
}{2}}X_{4}$ & $X_{5}$ & $e^{-\frac{3}{2}\varepsilon }X_{6}$ & $e^{-\frac{3}{2%
}\varepsilon }X_{7}$ & $X_{8}$ & $X_{9}$ \\ \hline\hline
\end{tabular}%
\label{tab2}%
\end{table}%

\subsection{Similarity transformations}

Among the elements of the one-dimensional Lie algebra not all the vector
fields reduce the number of the independent variables. For instance, the
symmetry vector $X^{1}=\partial _{H}$ cannot be used for the reduction of
the system of partial differential equations. However, these symmetries are
essential when the system reduces into a system of ordinary differential
equations.

\subsubsection{$\left\{ X^{3}\right\} $}

Application of the Lie symmetry vector $X^{3}$ leads to the reduced system $%
H\left( x,y\right) =h\left( y\right) ~,~W\left( x,y\right) =w\left( y\right)
$ in which $h_{yy}=0~,~w_{yy}=0.$ However such solution is not physically
accepted.

\subsubsection{$\left\{ X^{4}\right\} $}

From the symmetry vector $X^{4}$ it follows $H\left( x,y\right) =h\left(
x\right) ~,~W\left( x,y\right) =w\left( x\right) $ with reduced system~$%
h_{x}h_{xx}=0~,~\left( h_{x}w_{x}\right) _{x}=0$ with solutions~$\left\{
h\left( x\right) =h_{0}~,~w\left( x\right) \right\} $ and $\left\{ h\left(
x\right) =h_{0}\left( x-x_{1}\right) ~,~w\left( x\right) =w_{0}\left(
x-x_{2}\right) \right\} .~$The first solution is the Gibbons-Hawking
solution while the second solution is that of the maximal symmetric
spacetime of zero curvature, that is, it describes the flat space.

\subsubsection{$\left\{ X^{8}\right\} $}

Reduction with respect the Lie symmetry vector $X^{8}$ provides $H\left(
x,y\right) =h\left( x\right) y^{-2},~W\left( x,y\right) =w\left( x\right) $
where $h_{xx}h_{x}+6h=0$ ,~$\left( h_{x}w_{,x}\right) _{,x}=0$. Therefore, $%
\int \left( h_{0}-9h^{2}\right) ^{-\frac{1}{3}}dh=\left( x-x_{1}\right) $
and $w\left( x\right) =w_{0}+\int \frac{w_{1}}{h_{,x}}dx$. \ A special
solution is $h\left( x\right) =-\frac{1}{3}x^{3}$ with $w\left( x\right)
=w_{1}x^{-1}+w_{0}$. For the latter exact solution the background space
admits a four dimensional Lie algebra consisting of the generic vector field
$\left( c_{1}z+c_{2}\right) \partial _{t}+\left( c_{4}-4w_{1}c_{1}t+c_{1}%
\frac{2w_{1}}{x}\right) \partial _{y}+c_{3}\partial _{z}$.\thinspace\

Furthermore, for this exact solution the Ricciscalar of the spacetime is
found to be zero, that is, $R=0$, while the nonzero component of the
Einstein Tensor $G_{\mu \nu }=R_{\mu \nu }-\frac{1}{2}Rg_{\mu \nu }$\ is the
$G_{zz}=\frac{3}{\left( w_{1}\right) ^{2}}x^{4}$.

\subsubsection{$\left\{ X^{9}\right\} $}

From the Lie symmetry vector $X^{9}$ it follows the similarity
transformation $H\left( x,y\right) =h\left( \sigma \right) ~,~W\left(
x,y\right) =w\left( \sigma \right) $ in which $\sigma =yx^{-\frac{3}{2}}$and
$h\left( \sigma \right) ,~w\left( \sigma \right) $ satisfy the system $%
8h_{\sigma \sigma }-27\sigma ^{3}h_{\sigma }h_{\sigma \sigma }-45\sigma
^{2}\left( h_{\sigma }\right) ^{2}=0$~,~~$8w_{\sigma \sigma }-27\sigma
^{3}w_{\sigma }h_{\sigma \sigma }-90\sigma ^{2}h_{\sigma }w_{\sigma
}-27\sigma ^{3}h_{\sigma }w_{\sigma \sigma }=0$. The latter system can be
solved in terms of quadratures. However, a special solution of the system is
$h\left( \sigma \right) =-\frac{1}{3}\sigma ^{-2}~,~w\left( \sigma \right)
=w_{0}+w_{1}\sigma ^{\frac{2}{5}}$, that is, $H\left( x,y\right) =-\frac{1}{3%
}y^{-2}x^{3}$ , $W\left( x,y\right) =w_{0}+w_{1}y^{\frac{2}{5}}x^{-\frac{3}{5%
}}$. For that particular exact solution the Ricciscalar of the spacetime is
calculated $R=\frac{4}{3w_{1}}x^{\frac{8}{5}}y^{-\frac{12}{5}}$ while the
spacetime admits a three dimensional Killing algebra with generator $\left(
c_{1}t+c_{2}\right) \partial _{t}+3c_{1}\partial _{x}+2c_{1}y\partial
_{y}-\left( 2c_{1}z-c_{3}\right) \partial _{z}$. Finally the nonzero
components of the Einstein tensor are derived
\begin{equation*}
G_{tt}=\frac{244}{45}x^{2}y^{-4}~,~G_{ty}=-\frac{16}{15}xy^{-3}~,~G_{tz}=%
\frac{20}{9}x^{\frac{13}{5}}y^{-\frac{17}{5}}
\end{equation*}%
\begin{equation*}
G_{yz}=-\frac{2}{3w_{1}}x^{\frac{8}{5}}y^{-\frac{12}{5}}~,~G_{zz}=\frac{29}{9%
}x^{\frac{16}{5}}y^{-\frac{14}{5}}\text{.}
\end{equation*}

\subsubsection{$\left\{ a_{2}X_{2}+a_{8}X_{8}+a_{9}X_{9}\right\} $}

The generic symmetry vector $a_{2}X_{2}+a_{8}X_{8}+a_{9}X_{9}$ provides the
similarity transformation $H\left( x,y\right) =h\left( \xi \right) x^{\frac{%
a_{8}}{a_{9}}}\,,~W\left( x,y\right) =\frac{a_{2}}{a_{9}}\ln x+w\left( \xi
\right) $~,~$\xi =yx^{\frac{a_{8}-3a_{9}}{2a_{9}}}$. Hence, after the
application of the similarity transformation the gravitational field
equations are simplified in the following system%
\begin{eqnarray}
0 &=&h_{\xi \xi }\left( \left( a_{8}-3a_{9}\right) ^{3}h_{\xi }+8\left(
a_{9}\right) ^{3}+2a_{8}\left( a_{8}-3a_{9}\right) ^{2}\xi ^{2}h\right) +
\notag \\
&&+\left( a_{8}-a_{9}\right) \left( 4a_{8}h+5\left( a_{8}-3a_{9}\right) \xi
h_{\xi }\right) \left( 2a_{8}h+\left( a_{8}-3a_{9}\right) \xi h_{\xi
}\right) ,
\end{eqnarray}%
\begin{eqnarray}
0 &=&w_{\xi \xi }\left( \left( a_{8}-3a_{9}\right) ^{3}h_{\xi }+8\left(
a_{9}\right) ^{3}+2a_{8}\left( a_{8}-3a_{9}\right) ^{2}\xi ^{2}h\right) +
\notag \\
&&+\left( a_{8}-3a_{9}\right) ^{2}\xi ^{2}h_{\xi \xi }\left( 2a_{2}+\xi
\left( a_{8}-3a_{9}\right) h_{\xi }\right) +  \notag \\
&&+2\left( a_{8}-3a_{9}\right) \xi h_{\xi }\left( \left(
3a_{8}-5a_{9}\right) \left( a_{8}-3a_{9}\right) \xi w_{\xi }+a_{2}\left(
5a_{8}-7a_{9}\right) \right) +  \notag \\
&&+2a_{8}h\left( \xi \left( 3a_{8}-7a_{9}\right) \left( a_{8}-3a_{9}\right)
w_{\xi }+4a_{2}\left( a_{8}-2a_{9}\right) \right) .
\end{eqnarray}

The latter system can be solved by quadratures. Thus a closed-form solution
is
\begin{equation}
h\left( \xi \right) =-\frac{1}{3}\xi ^{-2}~,~w\left( \xi \right) =-w_{1}\xi
^{-\frac{2a_{9}}{a_{8}-5a_{9}}}-2\frac{a_{2}}{\left( a_{8}-a_{9}\right) }\ln
\xi +w_{0}~,~a_{8}\neq a_{9}.  \label{ss1}
\end{equation}

On the other hand in the special case where $a_{8}=a_{9}$ the closed-form
solution is
\begin{equation}
h\left( \xi \right) =h_{1}\xi +h_{0}~,~w\left( \xi \right) =\frac{1}{2}\frac{%
a_{2}}{a_{9}}\ln \left( 1+h_{0}\xi ^{2}\right) +\frac{w_{1}}{\sqrt{h_{0}}}%
\arctan \left( \sqrt{h_{0}}\xi \right) +w_{0}\text{.}  \label{ss2}
\end{equation}%
\

\begin{figure}[tbp]
\centering\includegraphics[width=1\textwidth]{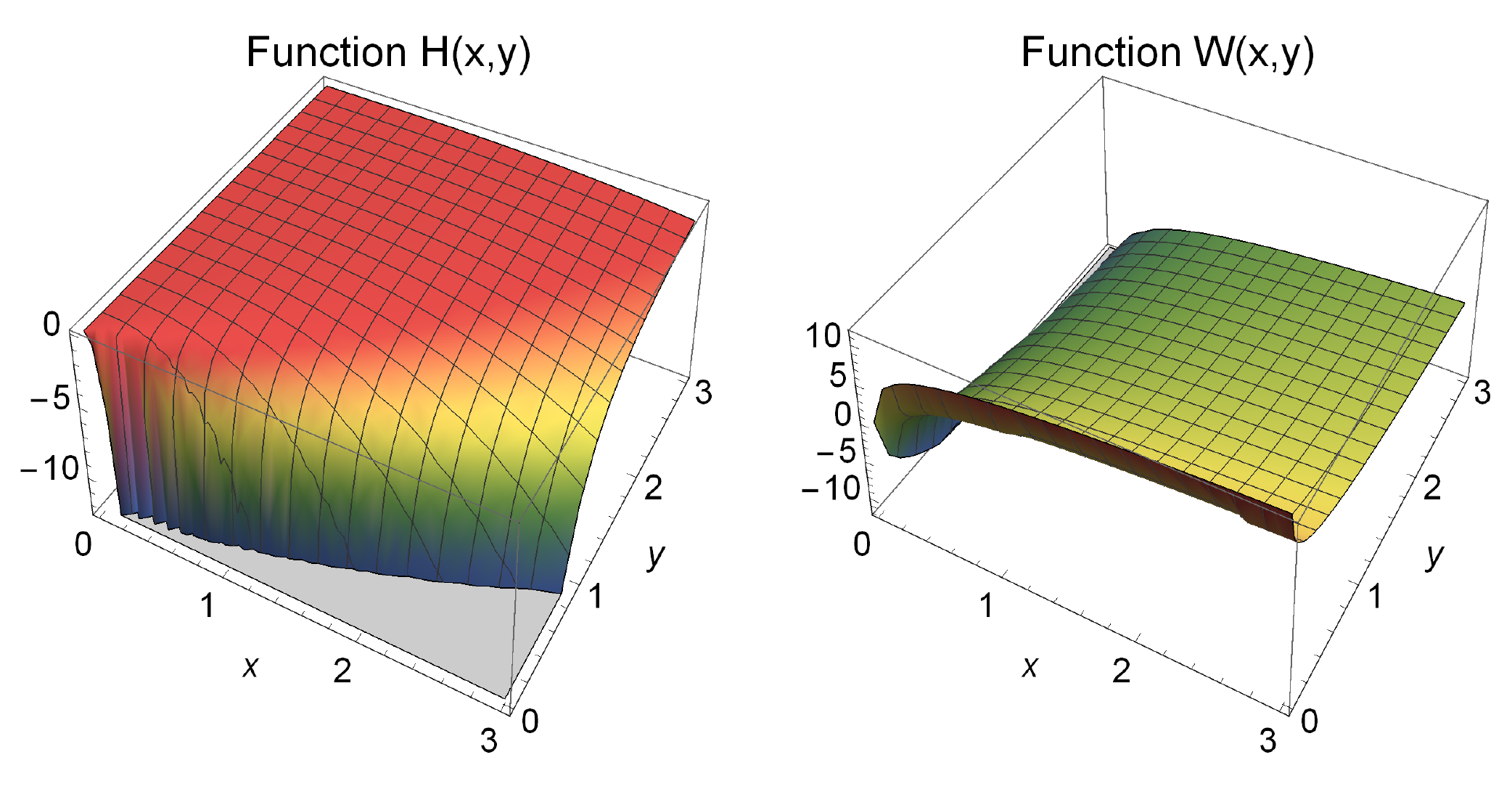}
\caption{Qualitative evolution for the similarity solution (\protect\ref{ss1}%
) of functions $H\left( x,y\right) $,~$W\left( x,y\right) $ provided by the
application of the symmetry vector $a_{2}X_{2}+a_{8}X_{8}+a_{9}X_{9}$. The
plots are for $\left( w_{1},a_{2},a_{8},a_{9}\right) =\left( 1,1,2,1\right) $%
. }
\label{mm1}
\end{figure}

\begin{figure}[tbp]
\centering\includegraphics[width=1\textwidth]{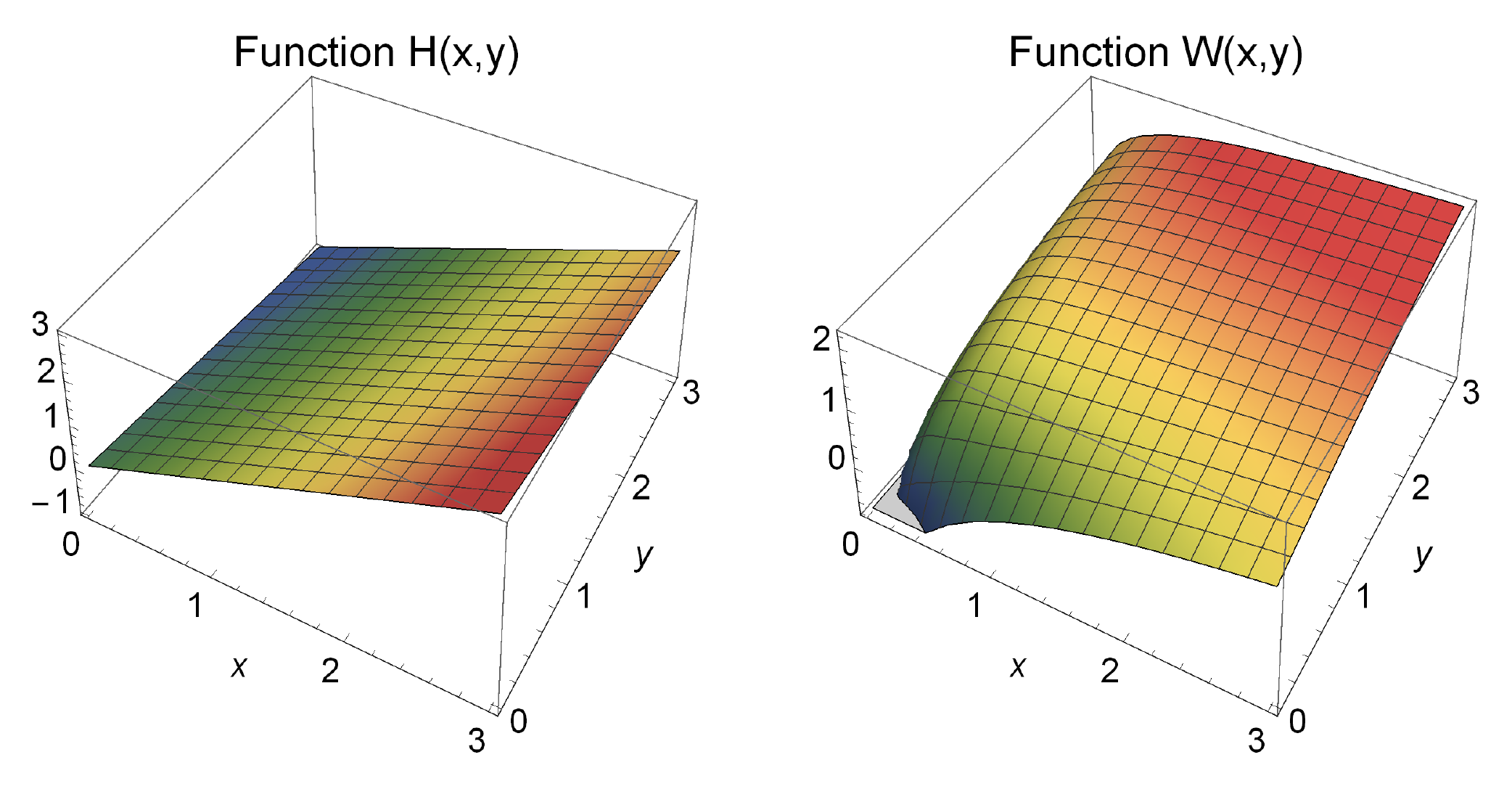}
\caption{Qualitative evolution for the similarity solution (\protect\ref{ss2}%
) of functions $H\left( x,y\right) $,~$W\left( x,y\right) $ provided by the
application of the symmetry vector $a_{2}X_{2}+a_{8}X_{8}+a_{8}X_{9}$. The
plots are for $\left( w_{1},a_{2},a_{8},h_{9}\right) =\left( 1,1,2,1\right) $%
. }
\label{mm2}
\end{figure}

For these specific solutions the spacetime is found to admit only two
isometries.\ In Figs. \ref{mm1} and \ref{mm2} we present the qualitative
evolution for the similarity solutions (\ref{ss1}), (\ref{ss2}).

\subsubsection{$\left\{ a_{3}X_{3}+a_{5}X_{5}+a_{8}X_{8}\right\} $}

From the vector field $a_{3}X_{3}+a_{5}X_{5}+a_{8}X_{8}$ it follows $H\left(
x,y\right) =e^{\frac{a_{8}}{a_{3}}x}h\left( \zeta \right) ~,~W\left(
x,y\right) =e^{\frac{a_{5}}{a_{3}}x}w\left( \zeta \right) $ ,~$\zeta =ye^{%
\frac{a_{8}}{2a_{3}}x}$. The reduced system is written as
\begin{equation}
0=h_{\zeta \zeta }\left( 8\left( a_{3}\right) ^{3}+\left( a_{8}\right)
^{3}\zeta ^{2}\left( \zeta h_{\zeta }+2h\right) \right) +\left( a_{8}\right)
^{3}\left( \zeta h_{\zeta }+2h\right) \left( 5\zeta h_{\zeta }+4h\right) ,
\end{equation}%
\begin{eqnarray}
0 &=&w_{\zeta \zeta }\left( 8\left( a_{3}\right) ^{3}+\left( a_{8}\right)
^{3}\zeta ^{2}\left( \zeta h_{\zeta }+2h\right) \right) +\left( a_{8}\right)
^{2}\zeta ^{2}h_{\zeta \zeta }\left( a_{8}\zeta w_{\zeta }+2a_{5}w\right) +
\notag \\
&&+a_{8}^{2}\zeta w_{\zeta }\left( \zeta \left( 3a_{8}+2a_{5}\right)
h_{\zeta }+\left( 3a_{8}+4a_{5}\right) h\right) +a_{5}a_{8}w\left( \zeta
\left( 2a_{5}+5a_{8}\right) h_{\zeta }+4\left( a_{5}+a_{8}\right) h\right) ,
\end{eqnarray}%
which can be solved in terms of quadratures.

\begin{figure}[tbp]
\centering\includegraphics[width=1\textwidth]{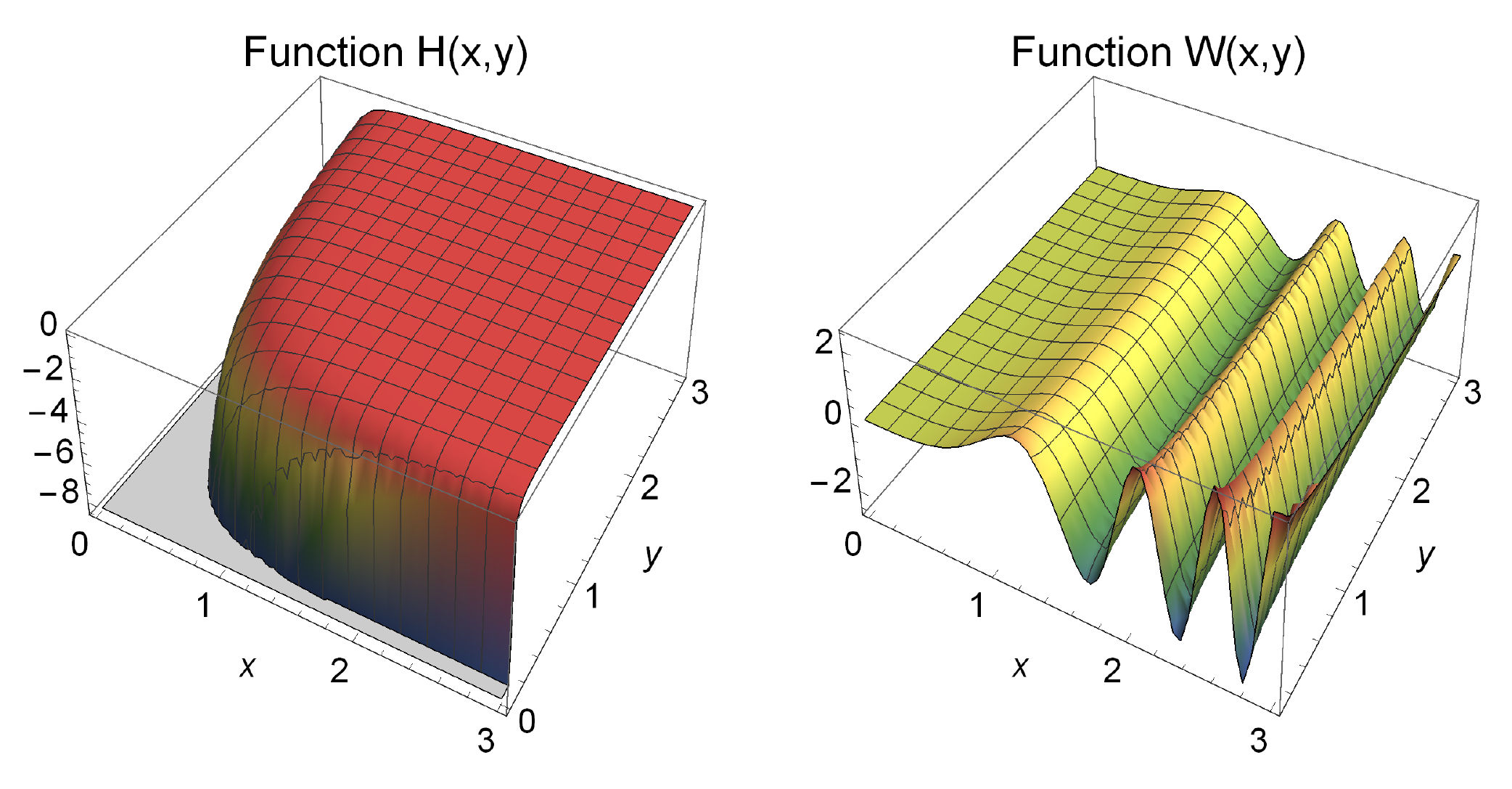}
\caption{Qualitative evolution for the similarity solution of functions $%
H\left( x,y\right) $,~$W\left( x,y\right) $ provided by the application of
the symmetry vector $a_{5}X_{5}+a_{8}X_{8}$. The plots are for $a_{5}=\frac{1%
}{2}~,~a_{8}=2,~w_{1}=1$ and $w_{2}=0.$}
\label{mm3}
\end{figure}

In the special case where $a_{3}=0$, the similarity transformation is
calculated $H\left( x,y\right) =h\left( x\right) y^{-2}$,~$W\left(
x,y\right) =w\left( x\right) y^{-\frac{2a_{5}}{a_{8}}}$ with reduced system $%
h_{xx}h_{x}+6h=0,~2a_{5}\left( 2a_{5}+a_{8}\right) w+a_{8}\left(
h_{x}w_{x}\right) _{,x}=0$. \ A\ closed-form solution of the latter system
is $h\left( x\right) =-\frac{1}{3}x^{3},~w\left( x\right) =w_{1}x^{2}J_{%
\frac{4}{5}}\left( \bar{a}x^{\frac{5}{2}}\right) +w_{2}x^{2}I_{\frac{4}{5}%
}\left( \bar{a}x^{\frac{5}{2}}\right) $, in which $\bar{a}=\frac{2\sqrt{%
3a_{5}\left( 2a_{5}+a_{8}\right) }}{a_{8}}$ and $J\left( x\right) ,~I\left(
x\right) $ are the Bessel functions. For the spacetime we find that it
admits a two dimensional Killing algebra. The qualitative evolution for the
latter similarity solution is given in Fig. \ref{mm3}.

\subsubsection{$\left\{ a_{1}X_{1}+a_{5}X_{5}+a_{9}X_{9}\right\} $}

We proceed with the application of the Lie symmetry vector $%
a_{1}X_{1}+a_{5}X_{5}+a_{9}X_{9}$. The corresponding similarity
transformation is $H\left( x,y\right) =\frac{a_{1}}{a_{9}}\ln x+h\left(
\sigma \right) ,~W\left( x,y\right) =w\left( \sigma \right) x^{\frac{a_{5}}{%
a_{9}}}$, where $\sigma =yx^{-\frac{3}{2}}$. The reduced system is%
\begin{equation}
0=a_{9}h_{\sigma \sigma }\left( 8a_{9}+18a_{1}\sigma ^{2}-27a_{9}\sigma
^{3}h_{\sigma }\right) -\left( 15a_{9}\sigma h_{\sigma }-4a_{1}\right)
\left( 3a_{9}\sigma h_{\sigma }-2a_{1}\right) ,
\end{equation}%
\begin{eqnarray}
0 &=&a_{9}^{2}w_{\sigma \sigma }\left( 8a_{9}-18a_{1}\sigma
^{2}-27a_{9}\sigma ^{3}h_{\sigma }\right) +9a_{9}^{2}\sigma ^{2}h_{\sigma
\sigma }\left( 2a_{5}h-3a_{9}\sigma w_{\sigma }\right) +  \notag \\
&&+6a_{9}\sigma w_{\sigma }\left( a_{1}\left( 7a_{9}-4a_{5}\right)
+3a_{9}\left( 2a_{5}-5a_{9}\right) \sigma h_{\sigma }\right) +  \notag \\
&&-6a_{5}a_{9}\left( 2a_{5}-7a_{9}\right) wh_{\sigma }+8a_{1}a_{5}\left(
a_{5}-2a_{9}\right) w.
\end{eqnarray}%
in which the solution can be expressed in terms of quadratures.

\begin{figure}[tbp]
\centering\includegraphics[width=1\textwidth]{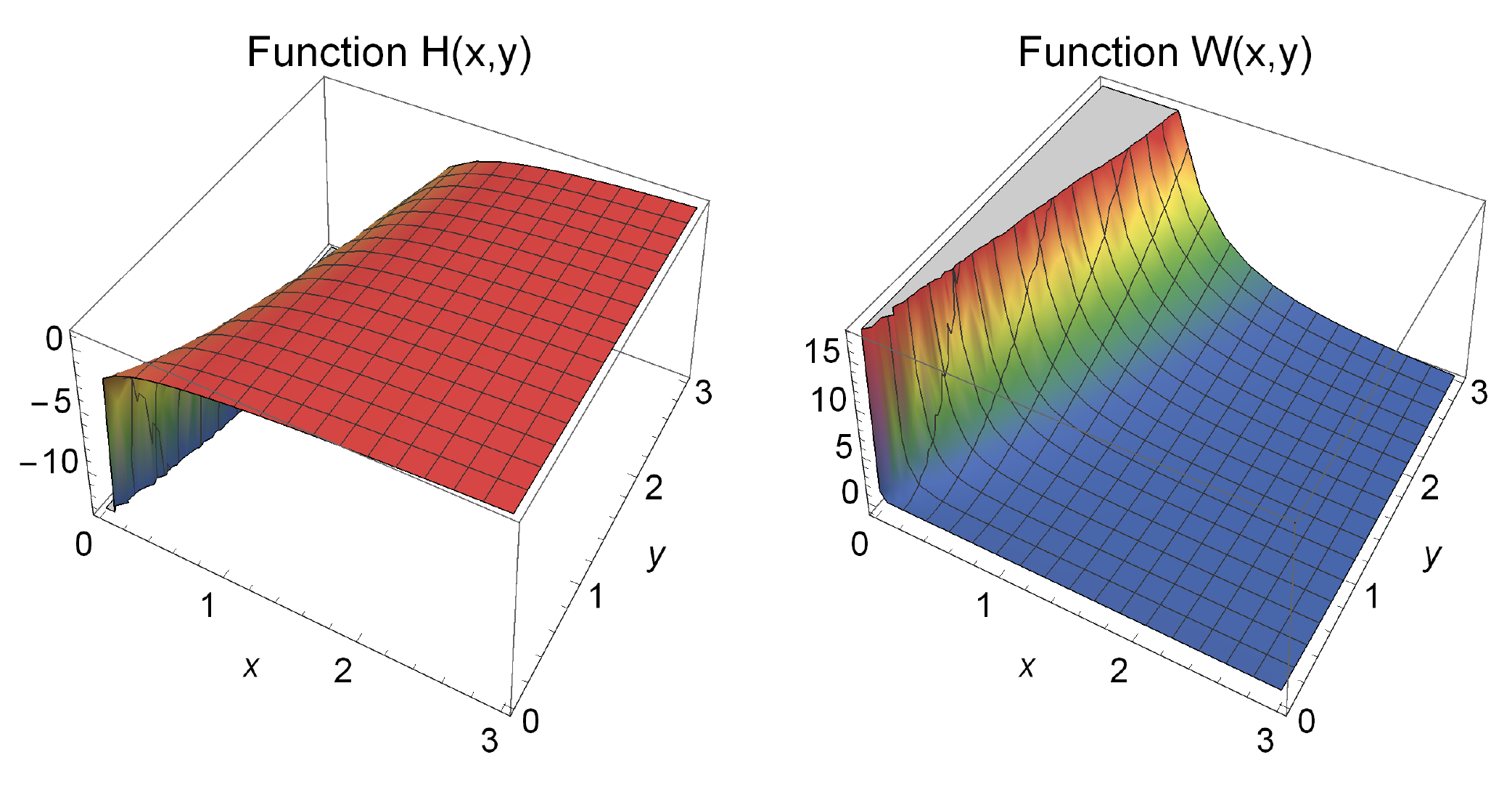}
\caption{Qualitative evolution for the similarity solution of functions $%
H\left( x,y\right) $,~$W\left( x,y\right) $ provided by the application of
the symmetry vector $a_{5}X_{5}+a_{9}X_{9}$. The plots are for $%
a_{5}=1~,~a_{8}=1,~w_{1}=1$ and $w_{2}=-1.$}
\label{mm4}
\end{figure}

On the other hand, for $a_{1}=0$ we find the closed-form solution $h\left(
\sigma \right) =-\frac{1}{3}\sigma ^{-2},~w\left( \sigma \right)
=w_{1}\sigma ^{\frac{2\left( a_{9}+a_{5}\right) }{5a_{9}}}+w_{2}\sigma ^{%
\frac{2a_{5}}{a_{9}}}$. For the latter exact solution the spacetime admits
two-dimensional Killing algebra, however when $w_{1}=0$, or $a_{5}+a_{9}=0$,
the spacetime admits the scaling symmetry $-\frac{2a_{5}}{2a_{5}+a_{9}}%
t\partial _{t}-\frac{2\left( a_{5}+a_{9}\right) }{2a_{5}+a_{9}}x\partial
_{x}-y\partial _{y}+z\partial _{z}$ as an additional isometry vector. The
qualitative evolution for the latter similarity solution is presented in
Fig. \ref{mm4}.

\subsubsection{$\left\{ a_{1}X_{1}+a_{2}X_{2}+a_{9}X_{9}\right\} $}

\begin{figure}[tbp]
\centering\includegraphics[width=1\textwidth]{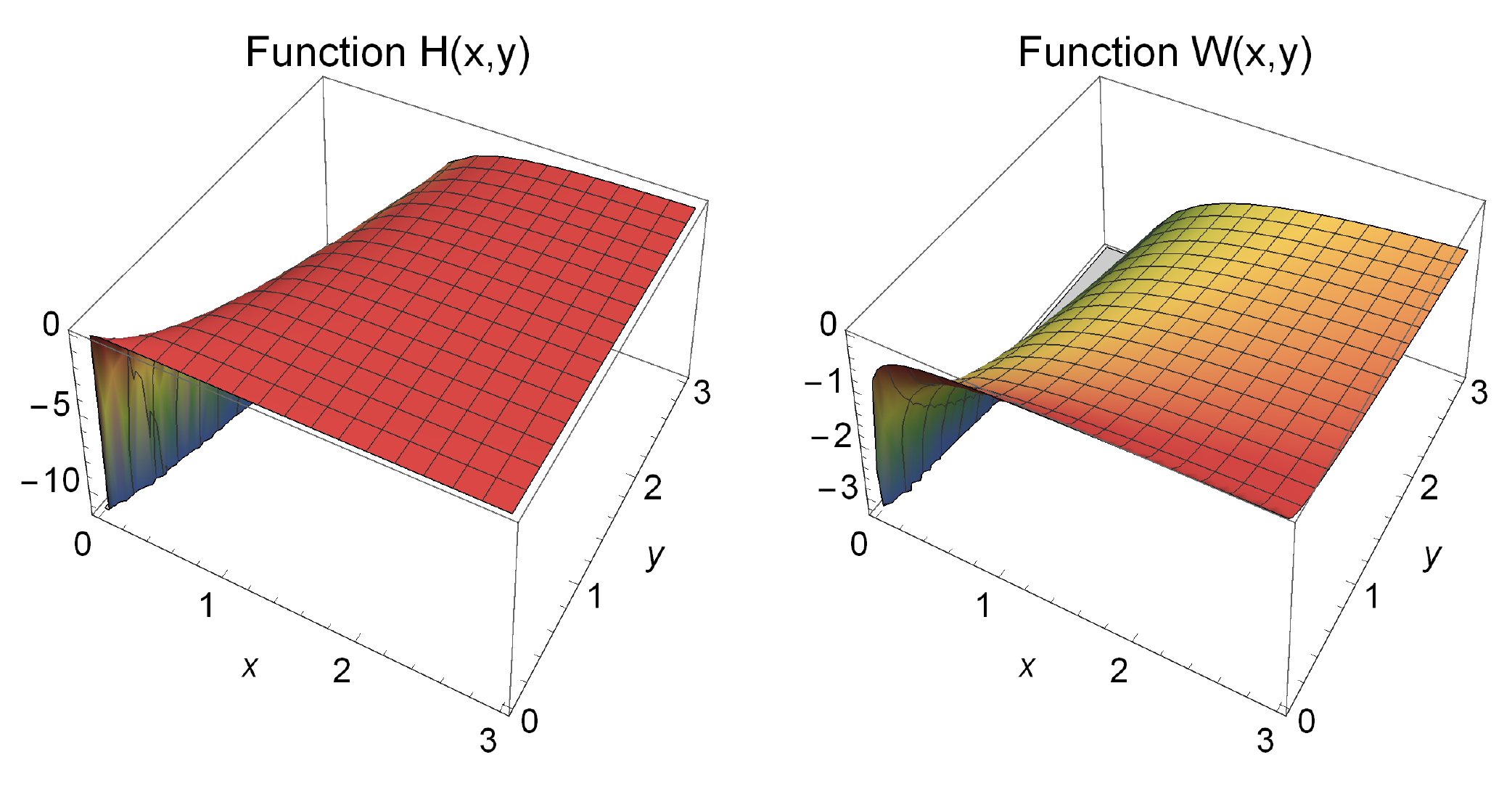}
\caption{Qualitative evolution for the similarity solution of functions $%
H\left( x,y\right) $,~$W\left( x,y\right) $ provided by the application of
the symmetry vector $a_{2}X_{2}+a_{9}X_{9}$. The plots are for $%
a_{2}=1,~a_{9}=1$.}
\label{mm5}
\end{figure}

From the Lie symmetry vector $a_{1}X_{1}+a_{2}X_{2}+a_{9}X_{9}$ it follows $%
H\left( x,y\right) =\frac{a_{1}}{a_{9}}\ln x+h\left( \sigma \right) ~,$ $%
W\left( x,y\right) =\frac{a_{2}}{a_{9}}\ln x+w\left( \sigma \right) $, $%
\sigma =yx^{-\frac{3}{2}}$. For the latter similarity transformation the
gravitational system is reduced to the following system of
ordinary-differential equations%
\begin{equation}
0=a_{9}h_{\sigma \sigma }\left( 18a_{1}\sigma ^{2}+8a_{9}-27a_{9}\sigma
^{3}h_{\sigma }\right) -2\left( 3a_{9}\sigma h_{gs}-2a_{1}\right) \left(
15a_{9}\sigma h_{\sigma }-4a_{1}\right) ,
\end{equation}

\begin{eqnarray}
0 &=&-a_{9}w_{\sigma \sigma }\left( 18a_{1}\sigma ^{2}+8a_{9}-27a_{9}\sigma
^{3}h_{\sigma }\right) +a_{8}\sigma ^{2}h_{\sigma \sigma }\left(
3a_{9}\sigma h_{\sigma }-2a_{2}\right) +  \notag \\
&&+42a_{9}\sigma \left( a_{1}w_{\sigma }+a_{2}h_{\sigma }\right)
-90a_{9}^{2}\sigma ^{2}h_{\sigma }-16a_{1}a_{2}.
\end{eqnarray}%
The later system can be solved by quadratures. However for $a_{1}=0$, we are
able to determine the exact solution~$h\left( \sigma \right) =-\frac{1}{3}%
\sigma ^{-2}~,~w\left( \sigma \right) =w_{1}\sigma ^{\frac{2}{5}}+2\frac{%
a_{2}}{a_{9}}\ln \sigma +w_{0}$. It easily follows that when $w_{1}a_{2}=0$,
a scaling symmetry similar to that found in the previous case exists as an
isometry for the gravitational background space.\ In Fig. \ref{mm5} we
present the evolution of the functions $H\left( x,y\right) $,~$W\left(
x,y\right) $\ for the similarity solution given by the symmetry vector $%
\left\{ a_{2}X_{2}+a_{9}X_{9}\right\} $.

\subsubsection{$\left\{ a_{2}X_{2}+a_{3}X_{3}+a_{8}X_{8}\right\} $}

In a similar way, from the Lie symmetry vector $%
a_{2}X_{2}+a_{3}X_{3}+a_{8}X_{8}$ it follows the similarity transformation $%
H\left( x,y\right) =e^{\frac{a_{8}}{a_{3}}x}h\left( \zeta \right) ~,~W\left(
x,y\right) =\frac{a_{2}}{a_{3}}x+w\left( \zeta \right) $ with reduced system%
\begin{equation}
0=h_{\zeta \zeta }\left( 8\left( a_{3}\right) ^{3}+\left( a_{8}\right)
^{3}\zeta ^{2}\left( 2h+\zeta h_{\zeta }\right) \right) +\left( a_{8}\right)
^{3}\left( 4h+5\zeta h_{\zeta }\right) \left( 2h+\zeta h_{\zeta }\right) ,
\end{equation}

\begin{eqnarray}
0 &=&w_{\zeta \zeta }\left( 8\left( a_{3}\right) ^{3}+\left( a_{8}\right)
^{3}\zeta ^{2}\left( 2h+\zeta g_{\zeta }\right) \right) +\left( a_{8}\right)
^{2}\zeta ^{2}h_{\zeta \zeta }\left( 2a_{2}+a_{8}\zeta h_{\zeta }\right) +
\notag \\
&&+2\left( a_{8}\right) ^{2}\left( 3a_{8}\zeta w_{\zeta }\left( \zeta
h_{\zeta }+h\right) +a_{2}\left( 5\zeta h_{\zeta }+4h\right) \right) .
\end{eqnarray}%
Again the solution can be written in terms of quadratures. However a
closed-form solution exists when $a_{3}=0$.

\begin{figure}[tbp]
\centering\includegraphics[width=1\textwidth]{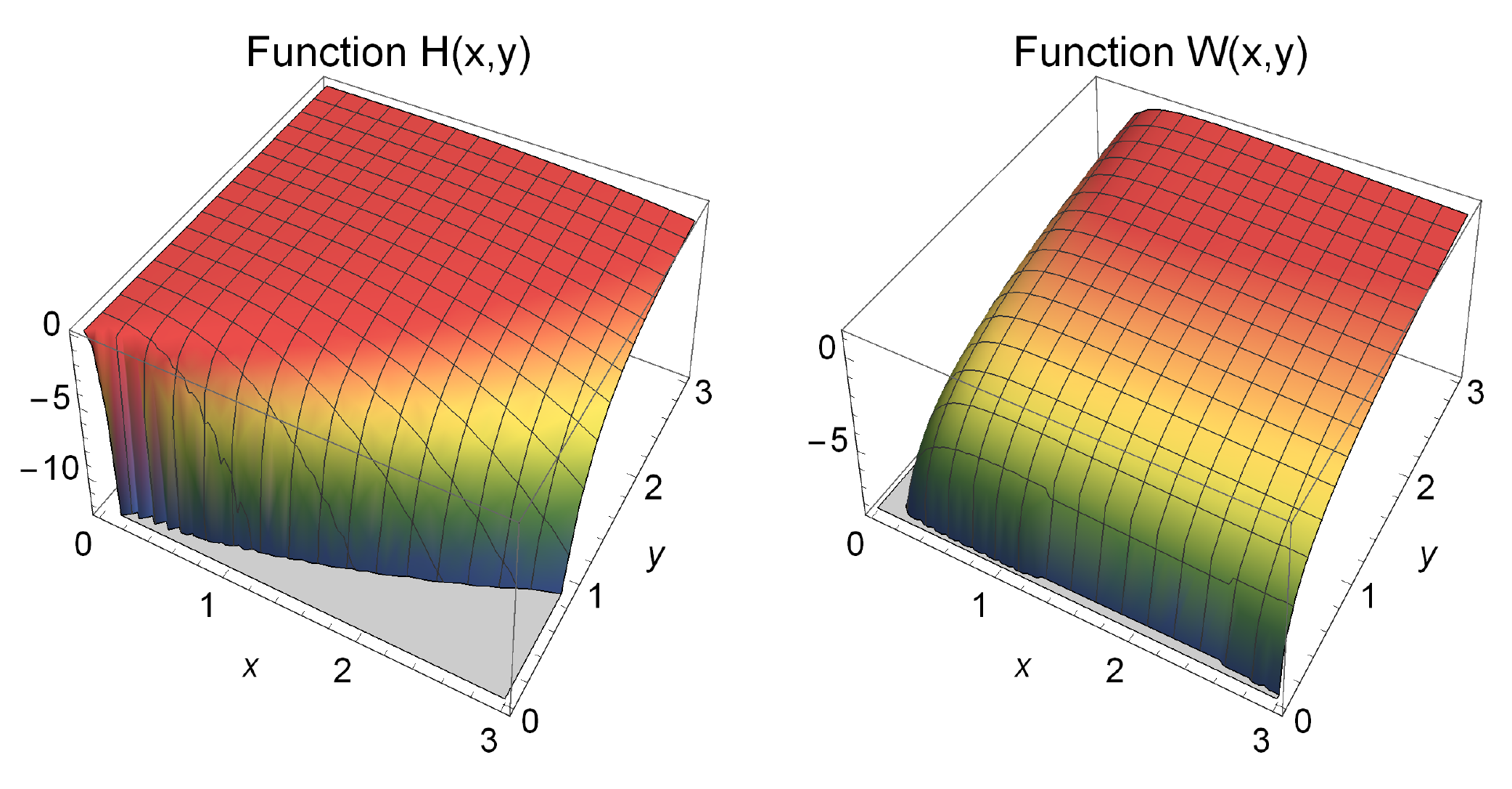}
\caption{Qualitative evolution for the similarity solution of functions $%
H\left( x,y\right) $,~$W\left( x,y\right) $ provided by the application of
the symmetry vector $a_{2}X_{2}+a_{8}X_{8}$. The plots are for $%
a_{2}=1,~a_{8}=-1$.}
\label{mm6}
\end{figure}

Indeed for $a_{3}=0$, the similarity transformation is $H\left( x,y\right)
=y^{-2}h\left( x\right) \,,~W\left( x,y\right) =-\frac{a_{2}}{a_{8}}\ln
y^{2}+w\left( x\right) $ with reduced system~$h_{xx}h_{x}+6h_{x}=0$, $%
2a_{2}+a_{8}\left( h_{x}w_{x}\right) _{x}=0$. The closed-form solution is $%
h\left( x\right) =-\frac{1}{3}x^{3},~\,w\left( x\right) =w_{1}x^{-1}+\frac{%
a_{2}}{a_{8}}\ln x^{2}+w_{0}$.\ The qualitative evolution for this
closed-form solution is presented in Fig. \ref{mm6}. Finally, for this exact
solution the spacetime admits always a two-dimensional Killing algebra.

\subsubsection{$\left\{ a_{3}X_{3}+a_{4}X_{4}+a_{5}X_{5}\right\} $}

From the generator $a_{3}X_{3}+a_{4}X_{4}+a_{5}X_{5}$, it follows $H\left(
x,y\right) =h\left( \mu \right) ,~W\left( x,y\right) =e^{\frac{a_{5}}{a_{3}}%
x}w\left( \mu \right) $ in which $\mu =y-\frac{a_{4}}{a_{3}}x$. The reduced
system is found to be~$h_{\mu \mu }\left( \left( a_{4}\right) ^{3}h_{\mu
}-\left( a_{3}\right) ^{3}\right) =0$, $w_{\mu \mu }\left( \left(
a_{4}\right) ^{3}h_{\mu }-\left( a_{3}\right) ^{3}\right) +\left(
a_{4}\right) ^{2}h_{\mu \mu }\left( a_{4}h_{\mu }-a_{5}h\right)
-a_{4}a_{5}h_{\zeta }\left( 2w_{\zeta }a_{4}-a_{5}w\right) =0$.

The analytic solution of the reduced system is given by the closed form
solution $h\left( \mu \right) =\left( \frac{a_{3}}{a_{4}}\right) ^{3}\mu
+h_{0}~,~w\left( \mu \right) =w_{1}e^{\frac{a_{5}}{2a_{4}}\mu }~$\ or $%
h\left( \mu \right) =h_{1}\mu +h_{0}$ ,~$w\left( \mu \right)
=w_{1}e^{M_{+}\mu }+w_{2}e^{M_{-}\mu }$ in which $M_{\pm }=a_{5}\frac{%
h_{1}a^{4}\pm a_{3}\sqrt{h_{1}a_{3}a_{4}}}{h_{1}a_{4}^{3}-a_{3}^{3}}$. \ For
this exact solution the spacetime admits an additional isometry which is $%
\frac{a_{3}}{a_{4}}\partial _{x}+\partial _{y}$.

For $a_{3}=0$, the similarity transformation is $H\left( x,y\right) =h\left(
x\right) ,~W\left( x,y\right) =e^{\frac{a_{5}}{a_{5}}y}w\left( x\right) $,
while the corresponding similarity solution is $h\left( x\right)
=h_{1}x+h_{0},~w\left( x\right) =w_{1}\sin \left( \frac{a_{5}}{a_{4}\sqrt{%
h_{1}}}x\right) +w_{2}\cos \left( \frac{a_{5}}{a_{4}\sqrt{h_{1}}}x\right) $.
In this case, the additional isometry is $\left( i\sqrt{h_{1}}\partial
_{x}+\partial _{y}\right) $, which means that the vector field is real when $%
h_{1}<0$.

\subsubsection{$\left\{ a_{3}X_{3}+a_{4}X_{4}+a_{6}X_{6}+a_{7}X_{7}\right\} $%
}

The vector field $a_{3}X_{3}+a_{4}X_{4}+a_{6}X_{6}+a_{7}X_{7}$ provides the
similarity transformation $H\left( x,y\right) =-\frac{a_{6}a_{4}}{2a_{3}^{2}}%
x^{2}+\frac{a_{6}}{a_{3}}xy+h\left( \mu \right) $ , $W\left( x,y\right) =-%
\frac{a_{7}a_{4}}{2a_{3}^{2}}x^{2}+\frac{a_{7}}{a_{3}}xy+w\left( \mu \right)
$ , $\mu =y-\frac{a_{4}}{a_{3}}x$. The reduced system is%
\begin{equation}
0=h_{\mu \mu }\left( \left( a_{4}\right) ^{3}h_{\mu }-\left( a_{3}\right)
^{3}-\left( a_{4}\right) ^{2}a_{6}\mu \right) +a_{4}a_{6}\left( a_{4}h_{\mu
}-a_{6}\mu \right) ,
\end{equation}%
\begin{eqnarray}
0 &=&w_{\mu \mu }\left( \left( a_{4}\right) ^{3}h_{\mu }-\left( a_{3}\right)
^{3}-\left( a_{4}\right) ^{2}a_{6}\mu \right) +\left( a_{4}\right)
^{2}h_{\mu \mu }\left( a_{4}w_{\mu }-a_{7}\mu \right) +  \notag \\
&&+a_{4}\left( a_{4}a_{6}w_{\mu }+a_{4}a_{7}h_{\mu }-2a_{6}a_{7}\mu \right) .
\end{eqnarray}

For the latter system we find $h\left( \mu \right) =\left( \frac{a_{3}}{a_{4}%
}\right) ^{3}\mu +\frac{a_{6}}{2a_{4}}\mu ^{2}+\frac{\left( a_{3}\right) ^{%
\frac{3}{2}}\left( \left( a_{3}\right) ^{3}+2a_{6}a_{4}^{2}\mu +h_{0}\right)
^{\frac{3}{2}}}{3\left( a_{4}\right) ^{5}a_{6}}~$, while $w\left( \mu
\right) $ is expressed in terms of quadratures. However for specific values
of the coefficient constants we can find closed-form solutions for both the
unknown functions.

When $a_{4}=0$, the similarity transformation reads $H\left( x,y\right) =%
\frac{a_{6}}{a_{3}}xy+h\left( y\right) $ , $W\left( x,y\right) =\frac{a_{7}}{%
a_{3}}xy+w\left( y\right) $ with the closed-form solution $h\left( y\right)
=h_{1}y+h_{0}~,~w\left( y\right) =w_{1}y+w_{0}$. \ For this exact solution
the spacetime admits an additional isometry, the vector fields $\frac{a_{3}}{%
2a_{7}}\partial _{x}+t\partial _{z}$.

For $a_{3}=0$, the similarity transformation reads $H\left( x,y\right) =%
\frac{a_{6}}{2a_{4}}y^{2}+h\left( x\right) ~,~W\left( x,y\right) =\frac{a_{7}%
}{2a_{4}}y^{2}+w\left( x\right) $ with the exact solution $h\left( x\right) =%
\frac{2\sqrt{2}}{3}\sqrt{-\frac{a_{6}}{a_{4}}}x^{\frac{3}{2}}~,~w\left(
x\right) =w_{1}\sqrt{x}-a_{7}\sqrt{-\frac{1}{a_{6}a_{4}}}x^{\frac{3}{2}%
}+w_{0}$. Hence, for this exact solution the spacetime admits the additional
isometry $\frac{a_{4}}{2a_{7}}\partial _{y}+t\partial _{z}$.

\subsubsection{$\left\{ a_{3}X_{3}+a_{4}X_{4}+a_{5}X_{5}+a_{6}X_{6}\right\} $%
}

From the Lie symmetry vector $a_{3}X_{3}+a_{4}X_{4}+a_{5}X_{5}+a_{6}X_{6}$
we find the similarity transformation $H\left( x,y\right) =-\frac{a_{6}a_{4}%
}{2a_{3}^{2}}x^{2}+\frac{a_{6}}{a_{3}}xy+h\left( \mu \right) $ , $W\left(
x,y\right) =w\left( \mu \right) e^{\frac{a_{5}}{a_{3}}x}~,~\mu =y-\frac{a_{4}%
}{a_{3}}x$ while the reduced system%
\begin{equation}
0=h_{\mu \mu }\left( \left( a_{4}\right) ^{3}h_{\mu }-\left( a_{3}\right)
^{3}-\left( a_{4}\right) ^{2}a_{6}\mu \right) +a_{4}a_{6}\left( a_{4}h_{\mu
}-a_{6}\mu \right) ,
\end{equation}%
\begin{eqnarray}
0 &=&w_{\mu \mu }\left( \left( a_{3}\right) ^{3}+\left( a_{4}\right)
^{2}a_{6}\mu -\left( a_{4}\right) ^{3}h_{\mu }\right) -\left( a_{4}\right)
^{2}h_{\mu \mu }\left( a_{4}w_{\mu }-a_{5}h\right) +  \notag \\
&&+a_{4}\left( w_{\mu }\left( a_{4}a_{6}-2a_{5}a_{6}\mu +2a_{4}a_{5}h_{\mu
}\right) -\left( a_{5}\right) ^{2}wh_{\mu }\right) -a_{5}a_{6}\left(
a_{4}-a_{5}\mu \right) w.
\end{eqnarray}%
which can be solved by quadratures.

\begin{figure}[tbp]
\centering\includegraphics[width=1\textwidth]{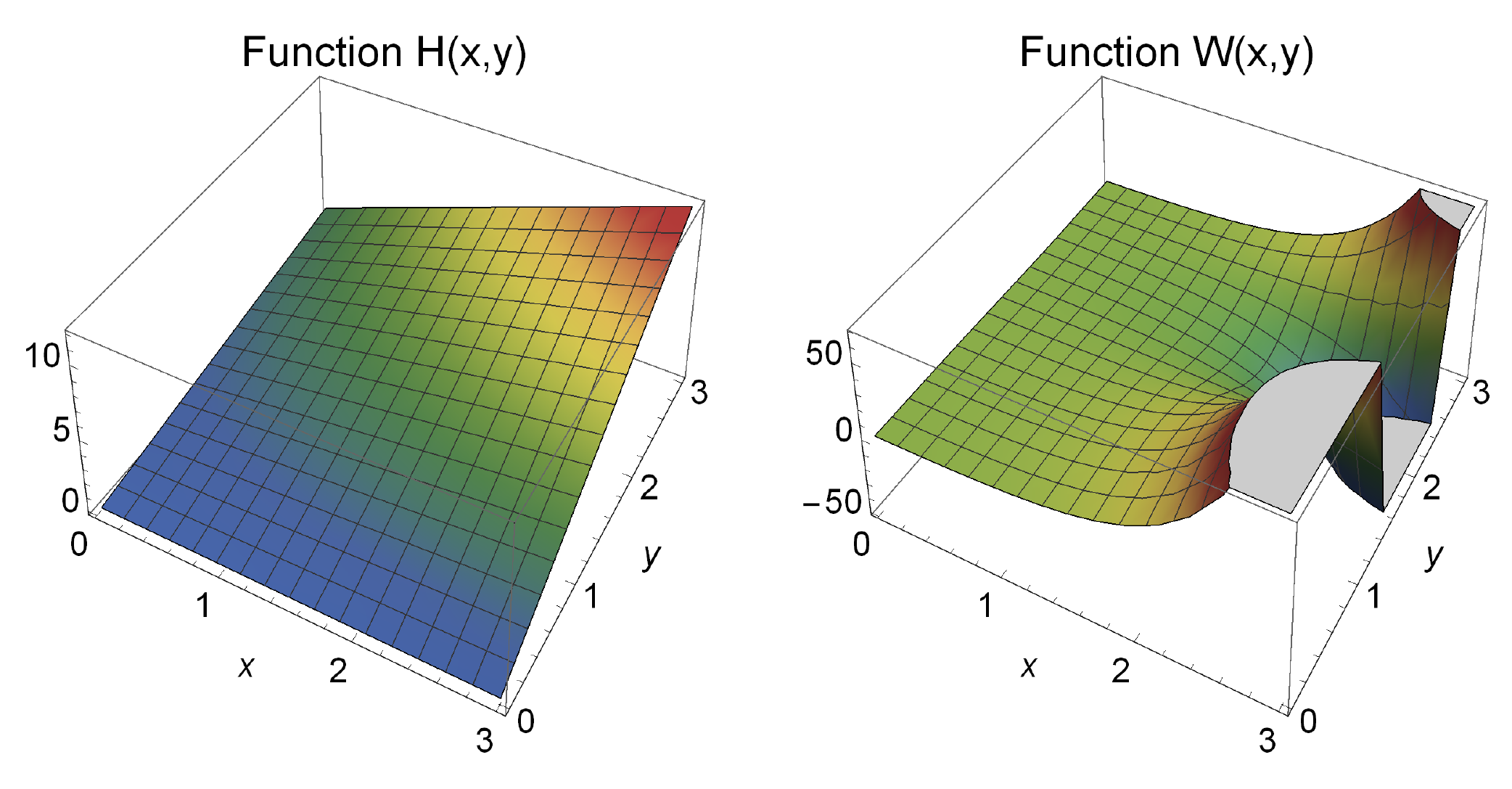}
\caption{Qualitative evolution for the similarity solution of functions $%
H\left( x,y\right) $,~$W\left( x,y\right) $ provided by the application of
the symmetry vector $a_{3}X_{3}+a_{5}X_{5}+a_{6}X_{6}$. The plots are for $%
a_{3}=1,~a_{5}=2,~a_{6}=1,~h_{1}=1$.}
\label{mm7}
\end{figure}

For $a_{4}$\thinspace $=0$, the closed form solution is expressed in terms
of the Airy functions, that is, $H\left( x,y\right) =\frac{a_{6}}{a_{3}}%
xy+h\left( y\right) $ , $W\left( x,y\right) =w\left( y\right) e^{\frac{a_{5}%
}{a_{3}}x}$, where $h\left( y\right) =h_{1}y+h_{0}$, $w\left( y\right)
=w_{1}Ai\left( \bar{a}y\right) +w_{2}Bi\left( \bar{a}y\right) $ in which $%
\bar{a}=-\left( a_{6}\left( a_{5}\right) ^{2}\left( a_{3}\right)
^{-3}\right) ^{\frac{1}{3}}$. The resulting spacetime admits a
two-dimensional Killing algebra.\ In Fig. \ref{mm7} the qualitative
evolution of the functions $H\left( x,y\right) $\ and $W\left( x,y\right) $\
is given.

On the other hand for $a_{3}=0$, we end with the closed form solution $%
H\left( x,y\right) =\frac{a_{6}}{2a_{4}}y^{2}+h\left( x\right) ~,~W\left(
x,y\right) =\frac{a_{6}}{2a_{4}}y^{2}+w\left( x\right) $ where $h\left(
x\right) =\frac{2\sqrt{2}}{3}\sqrt{-\frac{a_{6}}{a_{4}}}x^{\frac{3}{2}}$ and
$w\left( x\right) =w_{1}F_{2/3}\left( a^{\prime }x^{\frac{3}{2}}\right)
+w_{2}\sqrt{x}F_{4/3}\left( a^{\prime }x^{\frac{3}{2}}\right) $ where $F$ is
the hypergeometric function and $a=\frac{2\sqrt{2}\left( a_{5}\right) ^{2}}{9%
\sqrt{-a_{6}\left( a_{4}\right) ^{3}}}$. Again we find that the spacetime
admits a two-dimensional Killing algebra.

\subsubsection{$\left\{ a_{2}X_{2}+a_{3}X_{3}+a_{4}X_{4}+a_{6}X_{6}\right\} $%
}

From $a_{2}X_{2}+a_{3}X_{3}+a_{4}X_{4}+a_{6}X_{6}$ it follows $H\left(
x,y\right) =-\frac{a_{6}a_{4}}{2a_{3}^{2}}x^{2}+\frac{a_{6}}{a_{3}}%
xy+h\left( \mu \right) ~,W\left( x,y\right) =\frac{a_{2}}{a_{3}}x+w\left(
\mu \right) $, ~$\mu =y-\frac{a_{4}}{a_{3}}x$, \ with reduced system%
\begin{equation}
0=h_{\mu \mu }\left( \left( a_{3}\right) ^{3}-a_{6}\left( a_{4}\right)
^{2}\mu -\left( a_{4}\right) ^{3}h_{\mu }\right) +a_{4}a_{6}\left(
a_{4}h_{\mu }-\mu a_{6}\right) ,
\end{equation}%
\begin{equation}
0=w_{\mu \mu }\left( \left( a_{3}\right) ^{3}+\left( a_{4}\right)
^{2}a_{6}\mu -\left( a_{4}\right) ^{3}h_{\mu }\right) -\left( a_{4}w_{\mu
}-a_{2}\right) \left( \left( a_{4}\right) ^{2}h_{\mu \mu }-a_{4}a_{6}\right)
,
\end{equation}%
Therefore, the closed-form solution of the latter system is $h\left( \mu
\right) =\left( \frac{a_{3}}{a_{4}}\right) ^{3}\mu +\frac{a_{6}}{2a_{4}}\mu
^{2}+\frac{\left( a_{3}\right) ^{\frac{3}{2}}\left( \left( a_{3}\right)
^{3}+2a_{6}a_{4}^{2}\mu +h_{0}\right) ^{\frac{3}{2}}}{3\left( a_{4}\right)
^{5}a_{6}}~,~w\left( \mu \right) =w_{1}\sqrt{2a^{2}a_{6}\mu +\left(
a_{3}\right) ^{2}-h_{0}}+\frac{a_{2}}{a_{4}}\mu +w_{0}$.

In the special case with $a_{3}=0$, the similarity transformation is $%
H\left( x,y\right) =\frac{a_{6}}{2a_{4}}y^{2}+h\left( x\right) ,~W\left(
x,y\right) =\frac{a_{2}}{a_{4}}y+w\left( x\right) $ in which $h\left(
x\right) =\frac{2}{3}\sqrt{-2a_{4}a_{6}x^{3}}+h_{0}~,~w\left( x\right) =w_{1}%
\sqrt{x}+w_{0}\,$. In this case, the resulting spacetime admits as an
additional Killing symmetry the vector field $\partial _{y}$.

\subsubsection{$\left\{ a_{1}X_{1}+a_{2}X_{2}+a_{3}X_{3}+a_{6}X_{6}\right\} $%
}

From \ the vector field $a_{1}X_{1}+a_{2}X_{2}+a_{3}X_{3}+a_{6}X_{6}$ it
follows $H\left( x,y\right) =\frac{a_{1}x+a_{6}xy}{a_{3}}+h\left( y\right)
~,~W\left( x,y\right) =\frac{a_{2}}{a_{3}}x+w\left( y\right) $ in which $%
h_{yy}=0~$, $w_{yy}=0$, that is $h\left( y\right) =h_{1}y+h_{0}$ and $%
w\left( y\right) =w_{1}y+w_{0}$. In this case the spacetime is maximally
symmetric with zero Ricciscalar, that is, it becomes the flat space.

\subsubsection{$\left\{ a_{1}X_{1}+a_{3}X_{3}+a_{4}X_{4}+a_{5}X_{5}\right\} $%
}

Moreover, from the vector field $a_{1}X_{1}+a_{3}X_{3}+a_{4}X_{4}+a_{5}X_{5}$
we find the similarity transformation $H\left( x,y\right) =\frac{a_{1}}{a_{4}%
}y+h\left( x\right) ~,~W\left( x,y\right) =w\left( x\right) e^{\frac{a_{5}}{%
a_{4}}y}$, where the reduced system is $h_{x}h_{xx}=0,~a_{5}w+\left(
h_{x}w_{x}\right) _{x}=0$, with solution $h\left( x\right)
=h_{1}x+h_{0},~w\left( x\right) =w_{1}e^{i\frac{a_{5}}{a_{3}\sqrt{h_{1}}}%
x}+w_{2}e^{-i\frac{a_{5}}{a_{3}\sqrt{h_{1}}}x}$. For $w_{1}w_{2}=0$, the
spacetime admits a three dimensional Killing algebra. The third Killing
symmetry is the vector field $\sqrt{-h_{1}}\partial _{x}-\partial _{y}$ for $%
w_{2}=0$, or $\sqrt{-h_{1}}\partial _{x}+\partial _{y}$ for $w_{1}=0$.

\subsubsection{$\left\{ a_{1}X+a_{2}X_{2}+a_{3}X_{3}+a_{4}X_{4}\right\} $}

In the case where we apply the symmetry vector $%
a_{1}X+a_{2}X_{2}+a_{3}X_{3}+a_{4}X_{4}$ we end with the similarity solution
$H\left( x,y\right) =\frac{a_{1}}{a_{3}}x+h\left( \nu \right) ~,~W\left(
x,y\right) =w\left( \nu \right) ~$,~$\nu =y-2\frac{a_{4}}{a_{3}}x$; where $%
h\left( \nu \right) =h_{1}\nu +h_{0}~,~w\left( \nu \right) =w_{1}\nu +w_{0}$
or $h\left( \nu \right) =\frac{a_{3}^{3}+4a_{4}^{2}a_{1}}{8\left(
a_{4}\right) ^{3}}\nu +h_{0}$ with $w\left( \nu \right) $ arbitrary. In the
first solution the spacetime becomes maximally symmetric and specifically
the flat space, while in the case with arbitrary $w\left( \nu \right) $ the
background geometry admits as additional symmetry vector a translation
symmetry as in the previous case.

In this Section we investigate exact static ASD four-manifolds with at least
a $G_{2}$ Lie algebra. In Tables \ref{tab3} and \ref{tab4} we summarize all
the cases where the unknown functions are solved with the use of closed-form
expressions from the application of similarity transformations.

\begin{landscape}
\begin{table}[tbp] \centering%
\caption{Similarity transformations and exact solutions of static ASD
four-manifolds with at least a $G_2$ Lie algebra (1/2)}%
\begin{tabular}{ccc}
\hline\hline
\textbf{Symmetry vector} & \textbf{Similarity transformation} & \textbf{%
Exact Solution} \\ \hline
$X^{4}$ & $H=h\left( x\right) ,W=w\left( x\right) $ & $h\left( x\right)
=h_{0}~,~w\left( x\right) $ $;$ $h\left( x\right) =h_{0}\left(
x-x_{1}\right) ~,~w\left( x\right) =w_{0}\left( x-x_{2}\right) $ \\
$X^{8}$ & $H=h\left( x\right) y^{-2},~W=w\left( x\right) $ & $h\left(
x\right) =-\frac{1}{3}x^{3}~,~w\left( x\right) =w_{1}x^{-1}$ \\
$X^{9}$ & $H=h\left( \sigma \right) ~,~W=w\left( \sigma \right) ~,~\sigma
=yx^{-\frac{3}{2}}$ & $h\left( \sigma \right) =-\frac{1}{3}\sigma
^{-2}~,~w\left( \sigma \right) =w_{0}+w_{1}\sigma ^{\frac{2}{5}}$ \\
$a_{2}X_{2}+a_{8}X_{8}+a_{9}X_{9}$ & $H=h\left( \xi \right) x^{\frac{a_{8}}{%
a_{9}}}\,,~W=\frac{a_{2}}{a_{9}}\ln x+w\left( \xi \right) $~,~$\xi =yx^{%
\frac{a_{8}-3a_{9}}{2a_{9}}}$ & $h\left( \xi \right) =-\frac{1}{3}\xi
^{-2}~,~w\left( \xi \right) =-w_{1}\xi ^{-\frac{2a_{9}}{a_{8}-5a_{9}}}-2%
\frac{a_{2}}{\left( a_{8}-a_{9}\right) }\ln \xi $ \\
$a_{2}X_{2}+a_{9}X_{8}+a_{9}X_{9}$ & $H=h\left( \xi \right) x\,~,~W=\frac{%
a_{2}}{a_{9}}\ln x+w\left( \xi \right) $~,~$\xi =yx^{-1}$ & ${\small h}%
\left( \xi \right) {\small =h}_{1}{\small \xi +h}_{0}{\small ~,~w}\left( \xi
\right) {\small =}\frac{a_{2}}{2a_{9}}\ln \left( {\small 1+h}_{0}{\small \xi
}^{2}\right) {\small +}\frac{w_{1}}{\sqrt{h_{0}}}\arctan \left( \sqrt{h_{0}}%
{\small \xi }\right) $ \\
$a_{5}X_{5}+a_{8}X_{8}$ & $H=h\left( x\right) y^{-2}$,~$W=w\left( x\right)
y^{-\frac{2a_{5}}{a_{8}}}$ & $h=-\frac{1}{3}x^{3},~w=w_{1}x^{2}J_{\frac{4}{5}%
}\left( \bar{a}x^{\frac{5}{2}}\right) +w_{2}x^{2}I_{\frac{4}{5}}\left( \bar{a%
}x^{\frac{5}{2}}\right) $ \\
$a_{5}X_{5}+a_{9}X_{9}$ & $H=h\left( \sigma \right) ,~W=w\left( \sigma
\right) x^{\frac{a_{5}}{a_{9}}}~,~\sigma =yx^{-\frac{3}{2}}$ & $h\left(
\sigma \right) =-\frac{1}{3}\sigma ^{-2},~w\left( \sigma \right)
=w_{1}\sigma ^{\frac{2\left( a_{9}+a_{5}\right) }{5a_{9}}}+w_{2}\sigma ^{%
\frac{2a_{5}}{a_{9}}}$ \\
$a_{2}X_{2}+a_{9}X_{9}$ & $H=h\left( \sigma \right) ~,$ $W=\frac{a_{2}}{a_{9}%
}\ln x+w\left( \sigma \right) ~$, $\sigma =yx^{-\frac{3}{2}}$ & $h\left(
\sigma \right) =-\frac{1}{3}\sigma ^{-2}~,~w\left( \sigma \right)
=w_{1}\sigma ^{\frac{2}{5}}+2\frac{a_{2}}{a_{9}}\ln \sigma $ \\
$a_{2}X_{2}+a_{8}X_{8}$ & $H=y^{-2}h\left( x\right) \,,~W=-\frac{a_{2}}{a_{8}%
}\ln y^{2}+w\left( x\right) $ & $h\left( x\right) =-\frac{1}{3}%
x^{3},~\,w\left( x\right) =w_{1}x^{-1}+\frac{a_{2}}{a_{8}}\ln x^{2}$ \\
$a_{3}X_{3}+a_{4}X_{4}+a_{5}X_{5}$ & $H=h\left( \mu \right) ,~W=e^{\frac{%
a_{5}}{a_{3}}x}w\left( \mu \right) $ $,$ $\mu =y-\frac{a_{4}}{a_{3}}x$ & $%
{\small h=}\left( \frac{a_{3}}{a_{4}}\right) ^{3}{\small \mu +h}_{0}{\small %
,w=w}_{1}{\small e}^{\frac{a_{5}}{2a_{4}}\mu }~;{\small h=h}_{1}{\small \mu
+h}_{0}$,${\small w=w}_{1}{\small e}^{M_{+}\mu }{\small +w}_{2}{\small e}%
^{M_{-}\mu }$ \\
$a_{4}X_{4}+a_{5}X_{5}$ & $H=h\left( x\right) ,~W=e^{\frac{a_{5}}{a_{5}}%
y}w\left( x\right) $ & $h\left( x\right) =h_{1}x+h_{0},~w\left( x\right)
=w_{1}\sin \left( \frac{a_{5}}{a_{4}\sqrt{h_{1}}}x\right) +w_{2}\cos \left(
\frac{a_{5}}{a_{4}\sqrt{h_{1}}}x\right) $ \\ \hline\hline
\end{tabular}%
\label{tab3}%
\end{table}
\end{landscape}%

\begin{landscape}
\begin{table}[tbp] \centering%
\caption{Similarity transformations and exact solutions of static ASD
four-manifolds with at least a $G_2$ Lie algebra (2/2)}%
\begin{tabular}{ccc}
\hline\hline
\textbf{Symmetry vector} & \textbf{Similarity transformation} & \textbf{%
Exact Solution} \\
$a_{3}X_{3}+a_{6}X_{6}+a_{7}X_{7}$ & $H=\frac{a_{6}}{a_{3}}xy+h\left(
y\right) $ , $W=\frac{a_{7}}{a_{3}}xy+w\left( y\right) $ & $h\left( y\right)
=h_{1}y+h_{0}~,~w\left( y\right) =w_{1}y+w_{0}$. \\
$X_{4}+a_{6}X_{6}+a_{7}X_{7}$ & $H=\frac{a_{6}}{2a_{4}}y^{2}+h\left(
x\right) ~,~W=\frac{a_{7}}{2a_{4}}y^{2}+w\left( x\right) $ & $h\left(
x\right) =\frac{2\sqrt{2}}{3}\sqrt{-\frac{a_{6}}{a_{4}}}x^{\frac{3}{2}%
}~,~w\left( x\right) =w_{1}\sqrt{x}-a_{7}\sqrt{-\frac{1}{a_{6}a_{4}}}x^{%
\frac{3}{2}}$ \\
$a_{3}X_{3}+a_{5}X_{5}+a_{6}X_{6}$ & $H=\frac{a_{6}}{a_{3}}xy+h\left(
y\right) $ , $W=w\left( y\right) e^{\frac{a_{5}}{a_{3}}x}$ & $h\left(
y\right) =h_{1}y+h_{0}$, $w\left( y\right) =w_{1}Ai\left( \bar{a}y\right)
+w_{2}Bi\left( \bar{a}y\right) $ \\
$a_{4}X_{4}+a_{5}X_{5}+a_{6}X_{6}$ & $H=\frac{a_{6}}{2a_{4}}y^{2}+h\left(
x\right) ~,~W=\frac{a_{7}}{2a_{4}}y^{2}+w\left( x\right) $ & $h=\frac{2\sqrt{%
2}}{3}\sqrt{-\frac{a_{6}}{a_{4}}}x^{\frac{3}{2}}${\small ,}$%
w=w_{1}F_{2/3}\left( {\small a}^{\prime }{\small x}^{\frac{3}{2}}\right)
+w_{2}\sqrt{x}F_{4/3}\left( {\small a}^{\prime }{\small x}^{\frac{3}{2}%
}\right) $ \\
$a_{2}X_{2}+a_{3}X_{3}+a_{4}X_{4}+a_{6}X_{6}$ & $H=-\frac{a_{6}a_{4}}{%
2a_{3}^{2}}x^{2}+\frac{a_{6}}{a_{3}}xy+h\left( \mu \right) ~,$ & $h\left(
\mu \right) =\left( \frac{a_{3}}{a_{4}}\right) ^{3}\mu +\frac{a_{6}}{2a_{4}}%
\mu ^{2}+\frac{\left( a_{3}\right) ^{\frac{3}{2}}\left( \left( a_{3}\right)
^{3}+2a_{6}a_{4}^{2}\mu +h_{0}\right) ^{\frac{3}{2}}}{3\left( a_{4}\right)
^{5}a_{6}}~,~$ \\
& $W=\frac{a_{2}}{a_{3}}x+w\left( \mu \right) $, ~$\mu =y-\frac{a_{4}}{a_{3}}%
x$ & $w\left( \mu \right) =w_{1}\sqrt{2a^{2}a_{6}\mu +\left( a_{3}\right)
^{2}-h_{0}}+\frac{a_{2}}{a_{4}}\mu $ \\
$a_{2}X_{2}+a_{4}X_{4}+a_{6}X_{6}$ & $H=\frac{a_{6}}{2a_{4}}y^{2}+h\left(
x\right) ,~W=\frac{a_{2}}{a_{4}}y+w\left( x\right) $ & $h\left( x\right) =%
\frac{2}{3}\sqrt{-2a_{4}a_{6}x^{3}}+h_{0}~,~w\left( x\right) =w_{1}\sqrt{x}%
+w_{0}\,$ \\
$a_{1}X_{1}+a_{2}X_{2}+a_{3}X_{3}+a_{6}X_{6}$ & ${\small H}\left( x,y\right)
{\small =}\frac{a_{1}x+a_{6}xy}{a_{3}}{\small +h}\left( y\right) {\small ~,~W%
}\left( x,y\right) {\small =}\frac{a_{2}}{a_{3}}{\small x+w}\left( y\right) $
& $h\left( y\right) =h_{1}y+h_{0}$ , $w\left( y\right) =w_{1}y+w_{0}$ \\
$a_{1}X_{1}+a_{3}X_{3}+a_{4}X_{4}+a_{5}X_{5}$ & $H\left( x,y\right) =\frac{%
a_{1}}{a_{4}}y+h\left( x\right) ~,~W\left( x,y\right) =w\left( x\right) e^{%
\frac{a_{5}}{a_{4}}y}$ & $h\left( x\right) =h_{1}x+h_{0},~w\left( x\right)
=w_{1}e^{i\frac{a_{5}}{a_{3}\sqrt{h_{1}}}x}+w_{2}e^{-i\frac{a_{5}}{a_{3}%
\sqrt{h_{1}}}x}$ \\
$a_{1}X+a_{2}X_{2}+a_{3}X_{3}+a_{4}X_{4}$ & ${\small H}\left( x,y\right)
{\small =}\frac{a_{1}}{a_{3}}{\small x+h}\left( \nu \right) {\small ~,~W}%
\left( x,y\right) {\small =w}\left( \nu \right) {\small ~,~\mu =y-2}\frac{%
a_{4}}{a_{3}}{\small x}$ & $h=h_{1}\nu +h_{0}~,~w=w_{1}\nu ;$ $h=\frac{%
a_{3}^{3}+4a_{4}^{2}a_{1}}{8\left( a_{4}\right) ^{3}}\nu +h_{0}$~,$~w$
arbitrary \\ \hline\hline
\end{tabular}%
\label{tab4}%
\end{table}
\end{landscape}%

\section{Nonstatic spacetime}

\label{sec4}

In the case where the spacetime is nonstatic, the Lie point symmetries of
the system (\ref{as.01}), (\ref{as.02}) are
\begin{eqnarray*}
Y^{1} &=&\beta _{1}\left( t\right) \partial _{H}~,~Y^{2}=\beta _{2}\left(
t\right) \partial _{W}~,~Y^{3}=\left( \left( \beta _{3}\right) _{t}x+\frac{1%
}{2}\left( \beta _{3}\right) _{tt}\right) +\beta _{3}\left( t\right)
\partial _{x}~, \\
Y^{4} &=&\left( \frac{1}{2}\left( \beta _{4}\left( t\right) \right) _{tt}xy-%
\frac{1}{2}\left( \beta _{4}\left( t\right) \right) _{ttt}y^{3}\right) +%
\frac{1}{2}\left( \beta _{4}\left( t\right) \right) _{t}\partial _{x}+\beta
_{4}\left( t\right) \partial _{y}~~,~ \\
Y^{5} &=&W\partial _{W}~,~Y^{6}=\beta _{6}\left( t\right) y\partial
_{H}~,~Y^{7}=\beta _{7}\left( t\right) y\partial _{W}~,~
\end{eqnarray*}%
\begin{eqnarray*}
Y^{8} &=&\left( -\frac{1}{3}H\left( \beta _{8}\right) _{t}-\frac{1}{6}\left(
\left( \beta _{8}\right) _{tt}x^{2}-\left( \beta _{8}\right) _{ttt}xy^{4}-%
\frac{1}{12}\left( \beta _{4}\right) _{tttt}y^{4}\right) \right) \partial
_{H}-\frac{W}{3}\left( \beta _{8}\right) _{t}\partial _{W} \\
&&+\left( \frac{1}{3}x\left( \beta _{8}\right) _{t}+\frac{1}{6}y^{2}\left(
\beta _{8}\right) _{tt}\right) +\frac{2}{3}y\left( \beta _{8}\right)
_{t}\partial _{y}+\beta _{8}\left( t\right) \partial _{t}
\end{eqnarray*}%
\begin{equation*}
Y^{9}=\frac{2}{3}H\partial _{H}-\frac{1}{3}W\partial _{W}+\frac{1}{3}%
x\partial _{x}+\frac{1}{6}y\partial _{y}
\end{equation*}%
We observe that the gravitational system (\ref{as.01}), (\ref{as.02}) admits
an infinite Lie algebra. The latter infinite Lie algebra reduces to the
finite Lie algebra of the static case if we assume that all the coefficient
function $\beta _{8}\left( t\right) $ be constant and then apply the
similarity transformation which follows from $Y^{8}$, the reduced system is
that which was studied in the previous section. The infinite Lie symmetries
can be categorized in terms of nine families, where the eight families
describe infinite symmetries with arbitrary parameters the functions $%
\mathbf{\beta }\left( t\right) $.

The nonzero commutators of the admitted Lie algebra are
\begin{equation*}
\left[ Y^{1},Y^{8}\right] =-\frac{1}{3}\left( \beta _{1}\left( \beta
_{8}\right) _{t}+3\beta _{8}\left( \beta _{1}\right) _{t}\right) \partial
_{H}=-\frac{1}{3}\bar{Y}^{1}~~,~\left[ Y^{1},Y^{9}\right] =\frac{2}{3}Y^{1}~,
\end{equation*}%
\begin{equation*}
\left[ Y^{2},Y^{5}\right] =Y^{2}~,~\left[ Y^{2},Y^{8}\right] =-\frac{1}{3}%
\left( \beta _{2}\left( \beta _{8}\right) _{t}+3\beta _{8}\left( \beta
_{2}\right) _{t}\right) \partial _{W}=\bar{Y}^{2}~,~~\left[ Y^{2},Y^{9}%
\right] =-\frac{1}{3}Y^{2}~,
\end{equation*}%
\begin{equation*}
\left[ Y^{3},Y^{4}\right] =\frac{1}{2}\left( 2\beta _{4}\left( \beta
_{3}\right) _{t}-\beta _{3}\left( \beta _{4}\right) _{t}\right)
_{t}y\partial _{H}=\bar{Y}^{6}~,~\left[ Y^{3},Y^{8}\right] =-\frac{1}{3}\bar{%
Y}^{3}\left( 3\beta _{8}\left( \beta _{3}\right) _{t}-\beta _{3}\left( \beta
_{8}\right) _{,t}\right) ~,
\end{equation*}%
\begin{equation*}
\left[ Y^{3},Y^{9}\right] =\frac{1}{3}Y^{3}~,~\left[ Y^{4},Y^{6}\right]
=\beta _{4}Y^{6}=\bar{Y}^{6}~,~~\left[ Y^{4},Y^{7}\right] =\beta _{4}Y^{7}=%
\bar{Y}^{7}~,~
\end{equation*}%
\begin{equation*}
\left[ Y^{4},Y^{8}\right] =-\frac{1}{3}\bar{Y}^{8}\left( 3\beta _{8}\left(
\beta _{4}\right) _{t}-2\beta _{4}\left( \beta _{8}\right) _{t}\right) ~,~%
\left[ Y^{4},Y^{9}\right] =\frac{1}{6}Y^{4}~,~\left[ Y^{5},Y^{7}\right]
=-Y^{7},
\end{equation*}%
\begin{equation*}
\left[ Y^{6},Y^{8}\right] =\bar{Y}^{6}\left( \left( \beta _{8}\beta
_{6}\right) _{,t}\right) ~,~\left[ Y^{6},Y^{9}\right] =\frac{1}{2}Y^{6}~,~%
\left[ Y^{7},Y^{8}\right] =\bar{Y}^{7}\left( \left( \beta _{7}\beta
_{8}\right) _{,t}\right) ~\ ,~~\left[ Y^{7},Y^{9}\right] =-\frac{1}{2}Y^{7},
\end{equation*}%
and%
\begin{equation*}
\left[ Y^{8}\left( \beta _{8}\right) ,Y^{\prime 8}\left( \beta _{8}^{\prime
}\right) \right] =\bar{Y}^{8}\left( \beta _{8}\left( \beta _{8}^{\prime
}\right) _{,t}-\beta _{8}^{\prime }\left( \beta _{8}\right) _{,t}\right)
\end{equation*}

From the above result it follows that the gravitational system admits
infinite number of similarity transformations which can be used to determine
an infinite number of similarity transformations. Let us demonstrate the
application of the Lie symmetries to determine nonstatic ASD exact solutions.

Consider now the Lie symmetry vector $Y^{8}$ with $\beta _{8}\left( t\right)
=t$, that is%
\begin{equation*}
Y_{\beta _{8}\rightarrow t}^{8}=t\partial _{t}+\frac{1}{3}\left( x\partial
_{x}+2y\partial _{y}-H\partial _{H}-W\partial _{W}\right) .
\end{equation*}%
Application of the latter symmetry in the system (\ref{as.01}), (\ref{as.02}%
) \ provides the similarity transformation $H\left( t,x,y\right) =t^{-\frac{1%
}{3}}h\left( \chi ,\psi \right) ~,W\left( t,x,y\right) =w\left( \chi ,\psi
\right) $ in which $\chi =xt^{-\frac{1}{3}}$ and $\psi =yt^{-\frac{1}{3}}$.
The reduced equations for a system of two partial differential equations are%
\begin{equation}
3h_{\psi \psi }+\chi h_{\chi \chi }+2\psi h_{\psi \chi }+2h_{\chi }+3h_{\chi
}h_{\chi \chi }=0,  \label{sy.01}
\end{equation}%
\begin{equation}
3w_{\psi \psi }+\chi w_{\chi \chi }+2\psi w_{\psi \chi }+2w_{\chi }+3\left(
w_{\chi }h_{\chi }\right) =0.  \label{sy.02}
\end{equation}

The latter system consists a nine dimensional Lie algebra with elements
\begin{eqnarray*}
Y^{1\prime } &=&\partial _{H}~,~Y^{2\prime }=\partial _{W}~,~Y^{3^{\prime
}}=-\frac{1}{3}\left( \chi +\frac{1}{3}\psi ^{2}\right) +\partial _{\chi }~,
\\
Y^{4^{\prime }} &=&\left( \frac{1}{9}\chi \psi -\frac{2}{81}\psi ^{3}\right)
\partial _{H}+\frac{1}{3}\partial _{x}+\partial _{y}~~,~ \\
Y^{5} &=&W\partial _{W}~,~Y^{6^{\prime }}=Y\partial _{H}~,~Y^{7^{\prime
}}=Y\partial _{W}~,~ \\
Y^{8^{\prime }} &=&\left( -\frac{2}{9}\left( \chi ^{2}-\chi \psi ^{2}+\frac{1%
}{6}\psi ^{4}\right) \right) \partial _{H}+\left( \frac{2}{3}\chi +\frac{2}{9%
}\psi ^{2}\right) \partial _{\chi }+\psi \partial _{\psi }, \\
Y^{9^{\prime }} &=&H\partial _{H}+\frac{1}{2}x\partial _{x}+\frac{1}{4}%
\partial _{Y}
\end{eqnarray*}%
These vector fields form a finite Lie algebra which can be used for further
reduction as we did in the previous section. All these symmetry vectors are
reduced symmetries which follow from the infinite symmetries of the
time-dependent system. We demonstrate the application of the Lie symmetries
by present some exact solutions which follows from the application of the
latter similarity transformations.

From $Y^{9\prime }$ there follows the similarity solution $h\left( \chi
,\psi \right) =-\frac{1}{3}\chi \psi ^{2}~,~w\left( \chi ,\psi \right)
=\left( \frac{\psi ^{2}}{\chi }\right) ^{\frac{1}{3}}F\left( \frac{1}{2},%
\frac{2}{3};\frac{4}{3},\left( \frac{\psi ^{2}}{\chi }\right) \right) $ in
which $F$ is the hypergeometric function. For the latter solution the
spacetime is nonstatic. \ Moreover, from the symmetry vector $Y^{3^{\prime
}}+Y^{2^{\prime }}$ we find the similarity solution $h\left( \chi ,\psi
\right) =-\frac{\chi ^{2}}{6}+\frac{\chi \psi ^{2}}{9}-\frac{5}{324}\psi
^{4}+h_{1}\psi +h_{0}~,~w\left( \chi ,\psi \right) =\chi -\frac{1}{6}\psi
^{2}+w_{1}\psi +w_{0}$ which also provides a nonstatic spacetime. In a
similar way from the vector field $Y^{4\prime }$ there follows the
similarity transformation $h\left( \chi ,\psi \right) =-\frac{1}{6}\chi ^{2}+%
\frac{1}{9}\chi \psi ^{2}-\frac{5}{324}\psi ^{4}+h_{0}~,~w\left( \chi ,\psi
\right) =w\left( \psi ^{2}-6\chi \right) $ where $w$ is an arbitrary
function, or $h\left( \chi ,\psi \right) =-\frac{1}{6}\chi ^{2}+\frac{1}{9}%
\chi \psi ^{2}-\frac{5}{324}\psi ^{4}+h_{1}\left( \psi ^{2}-6\chi \right)
+h_{0}~,~w\left( \chi ,\psi \right) =w_{1}~\left( \psi ^{2}-6\chi \right)
+w_{0}$. Again for this solution the resulting spacetime admits only one
isometry vector, the field $\partial _{z}$.

\section{Conclusions}

\label{sec5}

In this work we investigated exact ASD four-dimensional null K\"{a}hler
manifolds by using the theory of Lie symmetries. In particular, we studied
the existence of one-parameter point transformations where the system (\ref%
{as.01}), (\ref{as.02}) remains invariant. In the general case we found that
there is an infinite number of Lie point symmetries, which belong to eight
families of infinite Lie symmetries which depend on eight time-dependent
coefficient variables. The infinite number of symmetries can be related with
the property that an ASD null K\"{a}hler manifold under a conformal
transformation remains an ASD manifold.

Furthermore, we studied the algebraic properties of the admitted Lie point
symmetries and we demonstrated the application of them on the determination
of exact solutions. We focused on the special case where the spacetime
admits a two dimensional Killing algebra and specifically when the spacetime
is static. In such specific case the admitted Lie point symmetries of the
reduced system are finite; they are nine. All these symmetries are reduced
symmetries from the infinite symmetries of the general case.

For the static case and for the finite Lie algebra we calculated the
commutators and the adjoint representation. The latter applied in order to
construct the one-dimensional optimal system, an essential tool for the
complete classification of the indepedent similarity transformations.
Finally, we use the invariants of the Lie point symmetries to reduce the
original system (\ref{as.01}), (\ref{as.02}) into a system of ordinary
differential equations which can be solved in all cases by quadratures.
Thus, we were able to determine exact closed-form solutions. For this exact
spacetimes we investigate the admitted isometries. For simplicity of the
presentation the results were presented in tables.

In a future work we plan to investigate the more general scenario of ASD
spacetimes without any isometry.

\end{document}